\documentclass[sigconf,screen]{acmart}
\AtBeginDocument{%
  \providecommand\BibTeX{{%
    \normalfont B\kern-0.5em{\scshape i\kern-0.25em b}\kern-0.8em\TeX}}}

\usepackage{siunitx}
\usepackage{mhchem}
\usepackage[ruled, linesnumbered, vlined, commentsnumbered]{algorithm2e}
\usepackage{xcolor,colortbl}
\usepackage{standalone}
\usepackage{tikz}
\usepackage{pgfplots}
\usetikzlibrary{pgfplots.statistics}
\usepackage{multirow}
\usepackage{subcaption} 
\usepackage{tcolorbox}
\usepackage{csquotes}
\usepackage{adjustbox}
\usepackage{balance}
\usepackage{lipsum}
\usepackage{pgf-pie}
\usepgfplotslibrary{statistics}
\usepgfplotslibrary{groupplots}
\usetikzlibrary{shadows}
\usepackage{amsmath}
\usepackage{amsfonts}
\usepackage{wrapfig}
\definecolor{beaublue}{rgb}{0.74, 0.83, 0.9}
\newtcolorbox{quotebox}{colback=white,boxrule=0.4pt,colframe=black,fonttitle=\bfseries,top=1pt,bottom=1pt,left=1pt,right=1pt}

\newtcolorbox{quotetitlebox}[1]{colback=beaublue,boxrule=0.4pt,colframe=black,fonttitle=\bfseries,top=1pt,bottom=1pt,left=1pt,right=1pt,title={#1}}
 \newcommand{\circo}{~\raisebox{1pt}{\tikz \draw[line width=0.3pt] circle(1.1pt);}}

\pgfplotsset{compat=1.11,
    /pgfplots/ybar legend/.style={
    /pgfplots/legend image code/.code={%
       \draw[##1,/tikz/.cd,yshift=-0.25em]
        (0cm,0cm) rectangle (3pt,0.8em);},
   },
}

\newcolumntype{P}[1]{>{\centering\arraybackslash}p{#1}}

\definecolor{one}{HTML}{2b7bba}
\definecolor{two}{HTML}{d52221}

\DeclareMathAlphabet\mathbfcal{OMS}{cmsy}{b}{n}
\SetKwInput{kwDeclare}{Declare}

\copyrightyear{2024}
\acmYear{2024}
\setcopyright{rightsretained}
\acmConference[ESEM '24]{Proceedings of the 18th ACM / IEEE International Symposium on Empirical Software Engineering and Measurement}{October 24--25, 2024}{Barcelona, Spain}
\acmBooktitle{Proceedings of the 18th ACM / IEEE International Symposium on Empirical Software Engineering and Measurement (ESEM '24), October 24--25, 2024, Barcelona, Spain}\acmDOI{10.1145/3674805.3686673}
\acmISBN{979-8-4007-1047-6/24/10}

\begin{document}


\title{Contexts Matter: An Empirical Study on Contextual Influence in Fairness Testing for Deep Learning Systems}
\author{Chengwen Du}
\affiliation{
  \institution{IDEAS Lab}
  \city{University of Birmingham\\}
  \country{United Kingdom}
}
\email{cxd394@student.bham.ac.uk}

\author{Tao Chen}
\authornote{Tao Chen is the corresponding author}
\affiliation{
  \institution{IDEAS Lab}
  \city{University of Birmingham\\}
  \country{United Kingdom}
}
\email{t.chen@bham.ac.uk}


\begin{abstract}

\textbf{Background:} Fairness testing for deep learning systems has been becoming increasingly important. However, much work assumes perfect context and conditions from the other parts: well-tuned hyperparameters for accuracy; rectified bias in data, and mitigated bias in the labeling. Yet, these are often difficult to achieve in practice due to their resource-/labour-intensive nature. \textbf{Aims:} In this paper, we aim to understand how varying contexts affect fairness testing outcomes. \textbf{Method:} We conduct an extensive empirical study, which covers $10,800$ cases, to investigate \textit{how} contexts can change the fairness testing result at the model level against the existing assumptions. We also study \textit{why} the outcomes were observed from the lens of correlation/fitness landscape analysis. \textbf{Results:} Our results show that different context types and settings generally lead to a significant impact on the testing, which is mainly caused by the shifts of the fitness landscape under varying contexts. \textbf{Conclusions:} Our findings provide key insights for practitioners to evaluate the test generators and hint at future research directions.



\end{abstract}



\maketitle

\section{Introduction}
\label{sec:introduciton}

Deep neural networks (DNNs) have become the fundamental part that underpins modern software systems, thanks to their ability to accurately learn complex non-linear correlations. The newly resulting type of software system, namely deep learning systems, has demonstrated remarkable success in many domains, such as healthcare, resource provisioning, criminal justice, and civil service~\cite{tramer2017fairtest,DBLP:journals/tsc/ChenB17,chan2018hiring,berk2021fairness}.

While accuracy and performance are undoubtedly important for deep learning systems, recent work has shown that these systems, when not engineered and built properly, can lead to severely unfair and biased results relating to some sensitive attributes such as gender, race, and age~\cite{DBLP:conf/sigsoft/BrunM18,DBLP:journals/corr/abs-2106-07849}. Often, such a fairness bug is not only overwhelming but can also lead to serious consequences: it has been reported that more than $75\%$ of the sentiment analysis tools provide higher sentiment intensity predictions for sentences associated with one race or one gender, exacerbating social inequity~\cite{DBLP:journals/expert/DuYZH21,DBLP:conf/starsem/KiritchenkoM18}.


Much work in software engineering has been proposed to discover such fairness bugs for the model embedded in deep learning systems~\cite{udeshi2018automated,zhang2020white,aggarwal2019black,galhotra2017fairness,aggarwal2019black,DBLP:conf/sigsoft/TaoSHF022,DBLP:conf/issta/ZhangZZ21,DBLP:conf/icse/ZhengCD0CJW0C22}. The paradigm, so-called fairness testing, has attracted increasing interest over the years. Like classic software testing for bugs, fairness testing automatically generates a set of test cases about data, such that they cause a deep learning system to make biased predictions, hence discovering fairness bugs~\cite{DBLP:journals/corr/abs-2207-10223,chen2023comprehensive}. This work focuses on testing at the model level, as it is one of the most active topics of research~\cite{DBLP:journals/corr/abs-2207-10223} and deep learning systems are model-centric in nature.

It is not difficult to envisage that the effectiveness of fairness testing at the model level involves some ``contexts'' set by the other parts of the deep learning systems (e.g., parameters and data)~\cite{DBLP:journals/corr/abs-2207-10223}: for example, how well the \textit{hyperparameters} have been tuned; the data ratio with respect to the sensitive attributes in the training data (i.e., \textit{selection bias}); and the sensitive attributes-related label ratio in the training samples (i.e., \textit{label bias}). However, existing work often ignores the presence of diverse contexts and their settings, instead, they assume the following situations in the evaluation:

\begin{itemize}
    \item There are well-optimized hyperparameters for accuracy (automatically tuned or derived from experience)~\cite{DBLP:journals/corr/abs-1905-05786,chakraborty2020making,gohar2023towards,tizpaz2022fairness}.
    \item The selection bias has been rectified~\cite{chen2022maat,chakraborty2020fairway,chakraborty2021bias,DBLP:conf/wacv/KarkkainenJ21,mambreyan2022dataset}.
    \item The training data is free of label bias or it has been fully dealt with in the labeling process~\cite{zhang2021ignorance,chakraborty2021bias,DBLP:conf/fairware-ws/ChakrabortyMT22,chakraborty2020fairway,DBLP:journals/corr/abs-2108-08504}.
\end{itemize}


As a result, current work relies on the belief that fairness testing at the model level runs under almost ``perfect'' conditions of the other parts in the deep learning systems. Yet, those assumptions are restricted and may not be desirable since tuning hyperparameters can be time-consuming~\cite{feurer2019hyperparameter} and the data collection (labeling) process can be expensive to change~\cite{DBLP:conf/iccv/ChenJ21,DBLP:conf/wacv/KarkkainenJ21}, hence it is highly possible that they are not properly handled before fairness testing for the purpose of rapid release in machine learning operations (MLOps). 


In fact, till now we still do not have a clear understanding of how these context settings can influence the test generators for fairness testing at the model level, and what are the causes thereof. Without such information, one can only believe the unjustified conjecture that some ``imperfect'' contexts would not considerably change the effectiveness of the proposed test generators from existing work. If such a conjecture is wrong, the practitioners may not be able to address the ``right'' problem in fairness testing or may even entail misleading evaluations and conclusions. This is the gap we intend to cover in this work via an empirical study.

In this paper, we conduct an extensive empirical study to understand the above. Our study covers 12 datasets for deep learning systems, three context types with 10 settings each, three test generators, and 10 instances of test adequacy and fairness metrics, leading to $10,800$ investigated cases. We do not only study \textit{how} contexts change the model level fairness testing result against the existing assumptions and the implication of distinct context settings, but also explain \textit{why} the phenomena were observed therein. 


In summary, we make the following contributions:

\begin{itemize}
    \item Extensive experiment and data analysis over $10,800$ cases on the role of contexts in fairness testing at the model level, in which the designs are derived from a systematic review. 
    \item We reveal that, compared with what is commonly assumed in existing work, (1) non-optimized hyperparameters often make a generator more struggle; while the presence of data bias boosts those generators. This can also affect the ranks of generators. (2) Changing the context settings generally leads to significant impacts on the fairness testing results.
    \item An in-depth study on why the context influences the outcome in the observations: they change the ruggedness with respect to local optima and/or the search guidance provided by the fairness testing landscape of test adequacy.
    \item We discover a weak monotonic correlation between the test adequacy and fairness metric under hyperparameter changes; for varying selection/label bias, such a correlation can be positive or negative based on the adequacy metric.
    
\end{itemize}

Drawing on the findings, we provide the following insights for testing the fairness of deep learning systems at the model level:

\begin{itemize}
    \item One must consider diverse settings of the contexts in the evaluation. Non-optimized hyperparameters and data bias can degrade and ameliorate existing generators, respectively.
    \item To improve test adequacy, current generators need to be strengthened with a strategy to handle ineffective fitness guidance under changing hyperparameters. In contrast, existing generators can focus on improving the speed of reaching better adequacy with varying selection/label bias.
    \item Better test adequacy does not necessarily lead to better fairness bug discovery, especially on metrics for the boundary of the neurons like NBC~\cite{zhou2020deepbillboard} and SNAC~\cite{ma2018deepgauge}.
\end{itemize}

All data, code, and materials are available at our repository: \texttt{\textcolor{blue}{\url{https://github.com/ideas-labo/FES}}}. The rest of this paper is organized as below: Section~\ref{sec:pre} introduces the preliminaries. Section~\ref{sec:methodology} elaborates our study protocol. Section~\ref{sec:results} discusses the results and Section~\ref{sec:exploration} explores the causes behind them. Section~\ref{sec:discussions} articulates the insights obtained.
Section~\ref{sec:tov},~\ref{sec:related}, and~\ref{sec:conclusion} present the threats to validity, related work, and conclusion, respectively.

\section{Preliminaries}
\label{sec:pre}


Here, we elaborate on the necessary background information and the scope we set for this work derived from our literature review.

\subsection{Literature Review}
\label{sec:slr}

To understand the trends in fairness testing and to design our study, we conducted a literature review on popular online repositories, i.e., ACM Library, IEEE Xplore, Google Scholar, ScienceDirect, Springer, and DBLP, using the following keywords:

\begingroup
\addtolength\leftmargini{-0.3cm}
\begin{quote}
    \textit{"software engineering" AND ("fairness testing" OR "fairness model testing") AND ("artificial intelligence" OR "neural network" OR "deep learning" OR "machine learning")}
\end{quote}
\endgroup

We focus on papers published in the past 5 years, which gives us 485 papers. After filtering duplicated studies, we temporarily selected a paper if it meets all the inclusion criteria below:

\begin{itemize}
    \item The paper aims to test or detect fairness bugs.
    \item The paper explicitly specifies the fairness metrics used.
    \item The paper at least mentions the possible context under which the fairness testing is conducted.
\end{itemize}

We then applied the following exclusion criteria to the previously included papers, which would be removed if they meet any:

\begin{itemize}
    \item Fairness testing is not part of the contributions to the work, but merely serves as a complementary component.
    \item The paper is a case study or empirical type of research.
    \item The paper has not been published in peer-reviewed venues.
\end{itemize}

The above leads to 53 relevant papers for us to analyze fairness testing for deep learning systems. These can be accessed at our repository: \texttt{\textcolor{blue}{\url{https://github.com/ideas-labo/FES/tree/main/papers}}}.


\begin{wrapfigure}[13]{r}{0.4\columnwidth}
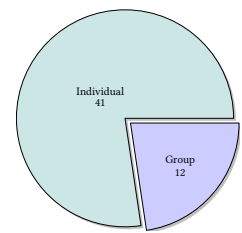

    \centering
    \includestandalone[width=0.35\columnwidth]{Figures/fairness}
    \caption{Distribution of the fairness categories.}
    \label{fig:fair}
\end{wrapfigure}

\subsection{Definition of Fairness}

We follow the definition of fairness bugs proposed by Chen \textit{et al.}~\cite{DBLP:journals/corr/abs-2207-10223}, which is ``\textit{A fairness bug refers to any imperfection in a software system that causes a discordance between the existing and required fairness conditions}''. Two main categories of fairness exist: \textit{individual fairness} and \textit{group fairness}~\cite{DBLP:conf/pkdd/KamishimaAAS12,DBLP:conf/kdd/SpeicherHGGSWZ18}. The former refers to the notion that similar individuals should be treated in a similar manner, in which the definition of ``similar'' also varies depending on the case. For example, in a loan approval system, two applicants with similar credit histories, income levels, and financial habits should be awarded the same loan amount or interest rate. The latter, group fairness, means that the deep learning system should give equal treatment to different populations or predefined groups. For example, in a candidate screening scenario, group fairness means that different demographic groups should have roughly equal acceptance rates. However, one criticism of group fairness is that it tends to ignore fine-grained unfair discrimination between individuals within groups~\cite{DBLP:conf/aies/Fleisher21}.



Figure~\ref{fig:fair} shows the proportions in the 53 identified papers that focus on either of the two categories of fairness. Clearly, individual fairness is at least $3\times$ more common than its group counterpart. Notably, nearly all papers that propose new fairness testing generators at the model level consider individual fairness~\cite{udeshi2018automated,zhang2020white,galhotra2017fairness,aggarwal2019black,DBLP:conf/sigsoft/TaoSHF022,DBLP:conf/issta/ZhangZZ21,DBLP:conf/icse/ZhengCD0CJW0C22}. Therefore, we focus on individual fairness in this work.

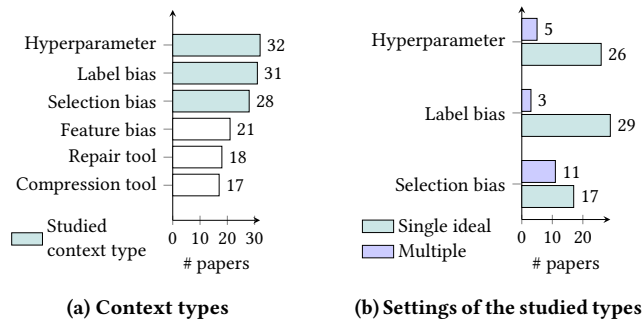
\begin{figure}
     \centering
     \begin{subfigure}[t]{0.45\columnwidth}
          \centering
          \begin{tikzpicture}
\begin{axis}[
    axis x line  = bottom,
    axis y line  = left  ,
    xbar=-0.35cm, 
    enlarge y limits=0.25,
    symbolic y coords={Compression tool,Repair tool,Feature bias,Selection bias,Label bias,Hyperparameter},
    ytick={Compression tool,Repair tool,Feature bias,Selection bias,Label bias,Hyperparameter},
    xlabel=\# papers,
    legend style={at={(-1.05,0.05)},cells={align=left},draw=none,anchor=north,legend columns=1},
    legend cell align={left},
    y tick label style={rotate=0,anchor=east},
    nodes near coords,
    height=5cm,
    width=3cm,
    xmin=0,
    legend image code/.code={
        \draw [#1] (-0.1cm,-0.1cm) rectangle (0.4cm,0.1cm); },
    ]

\addplot [fill=teal!20,draw=black] coordinates {(32,Hyperparameter)};
\addplot [fill=teal!20,draw=black] coordinates {(31,Label bias)};
\addplot [fill=teal!20,draw=black] coordinates {(28,Selection bias)};
\addplot [fill=white,draw=black] coordinates {(21,Feature bias)};
\addplot [fill=white,draw=black] coordinates {(18,Repair tool)};
\addplot [fill=white,draw=black] coordinates {(17,Compression tool)};


\addlegendentry{Studied \\context type}
\end{axis}
\end{tikzpicture}
          \subcaption{Context types}
     \end{subfigure}
     \hfill
     \begin{subfigure}[t]{0.45\columnwidth}
          \centering
          \begin{tikzpicture}
\begin{axis}[
    axis x line  = bottom,
    axis y line  = left  ,
    xbar=0.05cm, 
    enlarge y limits=0.25,
    symbolic y coords={Selection bias,Label bias,Hyperparameter},
    ytick={Selection bias,Label bias,Hyperparameter},
    xlabel=\# papers,
    legend style={at={(-1.05,0.05)},cells={align=left},draw=none,anchor=north,legend columns=1},
    legend cell align={left},
    y tick label style={rotate=0,anchor=east},
    nodes near coords,
    height=5cm,
    width=3cm,
    xmin=0,
     legend image code/.code={
        \draw [#1] (-0.1cm,-0.1cm) rectangle (0.4cm,0.1cm); },
    ]

\addplot [fill=teal!20,draw=black] coordinates {(26,{Hyperparameter}) (29,{Label bias}) (17,{Selection bias})};
\addplot [fill=blue!20,draw=black] coordinates {(5,{Hyperparameter}) (3,{Label bias}) (11,{Selection bias})};


\addlegendentry{Single ideal}
\addlegendentry{Multiple}
\end{axis}
\end{tikzpicture}
          \subcaption{Settings of the studied types}
     \end{subfigure}
    \caption{Distribution of context type the setting considered.}
    \label{fig:context}
\end{figure}

\subsection{Fairness Testing}

Fairness testing refers to the process designed to reveal fairness bugs through code execution in a deep learning system~\cite{DBLP:journals/corr/abs-2207-10223}. At the model level, the test generator mutates input data to generate sufficiently diverse and discriminatory samples~\cite{DBLP:conf/cvpr/SzegedyVISW16}. Since a deep learning system is model-centric, testing at the model is a key topic for fairness testing~\cite{DBLP:journals/corr/abs-2207-10223,chen2023comprehensive}, which is also our focus in this paper.



Like classic software testing, testing the properties such as fairness of deep learning systems can distinguish the notion of test adequacy and fairness metrics, where the former determines when the testing should stop while the latter measures how well the test generators perform in terms of finding fairness bugs~\cite{DBLP:journals/corr/abs-2207-10223}. Although the relationships between those types of metrics are unclear, recent studies have indicated that pairing general test adequacy metrics with fairness testing generators is a promising solution~\cite{DBLP:journals/tosem/ChenWMSSZC23,DBLP:journals/corr/abs-2207-10223}.


\subsection{Contexts in Fairness Testing at Model Level}



Although we aim to test fairness bugs at the model level, other parts involved in a deep learning system can form different aspects of the contexts. For example, the distribution of the training data samples with respect to the sensitive attributes and the hyperparameter configuration. In this work, \textbf{\textit{context type}} refers to the category of condition from the other parts in a deep learning system, e.g., hyperparameters and label bias. In contrast, we use \textbf{\textit{context setting}} to denote certain conditions of a context type, e.g., a configuration of hyperparameters and a label ratio in the training data.

To confirm what context types are explicitly considered and to understand their nature in fairness testing for deep learning systems, we extracted information from the 53 papers identified. From Figure~\ref{fig:context}a, it is clear that the following three context types are more prevalent, which are our focuses in this work:

\begin{itemize}
    \item \textbf{Hyperparameters, e.g., ~\cite{DBLP:journals/corr/abs-1905-05786,DBLP:conf/kdd/Corbett-DaviesP17,DBLP:journals/tse/Nair0MSA20,DBLP:journals/corr/abs-2011-11001}} refer to the key parameters that can influence the model's behaviors, such as learning rates and the number of neurons/layers. 
    They are often assumed to be well-tuned for better model accuracy.

    \item \textbf{Selection bias, e.g., ~\cite{DBLP:conf/fat/YeomT21,DBLP:conf/fairware-ws/ChakrabortyMT22,DBLP:conf/wacv/KarkkainenJ21}} indicates the biases that arise during the sampling process for collecting the data. This would often introduce an unexpected correlation between sensitive attributes and the outcome. For example, the \textsc{Compas} dataset~\cite{compasdataset2016} has been shown to exhibit unintended correlations between race and recidivism~\cite{DBLP:conf/nips/WickpT19} as it was collected during a specific time period (2013 to 2014) and from a particular county in Florida. As such, the inherent policing patterns make it susceptible to unintentional correlations.

    \item \textbf{Label bias, e.g., ~\cite{DBLP:journals/corr/abs-2110-01109,DBLP:journals/corr/abs-2108-08504,compasdataset2016}} refers to the biased/unfair outcome labels in the training data as a result of the manual labeling process. This can arise due to historical or societal biases; flawed data collection processes; or the subjective nature of label assignments conducted by humans/algorithms.
\end{itemize}

For those context types, Figure~\ref{fig:context}b further summarizes the number of context settings considered for the evaluation therein. We see that for the majority of the cases, a perfect situation is assumed: the optimized hyperparameters; removed/rectified selection bias; or omitted/mitigated label bias. Notably, $77\%$ (41/53) papers evaluate the fairness test generators with a specific assumption for all three context types; all papers rely on the fixed context setting for at least one context type. Since it is expensive to reach those perfect states in practice, the above motivates our study: can we really ignore the diverse contexts for testing the fairness of deep learning systems?

\section{Methodology}
\label{sec:methodology}

In what follows, we describe the methodology of our study. 


\subsection{Initial Research Questions}

We study two initial research questions (RQ$_1$ and RQ$_2$) to understand the role of contexts (i.e., hyperparameters, selection bias, and label bias) in fairness testing of deep learning systems:



\begin{quotebox}
   \noindent
   \textit{\textbf{RQ$_1$:} What implications do the contexts bring against the current assumptions in the existing work for fairness testing?}
\end{quotebox}

Understanding RQ$_1$ would strengthen the basic motivation of this work: do we need to consider the fact that the context may not always be the same as the perfect/ideal conditions assumed in existing work? Indeed, if different contexts pose insignificant change to the testing results then there is no need to explicitly take them into account in the evaluation of fairness testing. 


Answering the above is not straightforward, because there are simply too many settings for the diverse context types, e.g., the possible hyperparameter configurations, the percentage of biased data with respect to the sensitive attributes, and the ratio on the biased labels. This motivates us to study another interrelated RQ:

\begin{quotebox}
   \noindent
   \textit{\textbf{RQ$_2$:} To what extent do the context settings influence the fairness testing outcomes between each other?}
\end{quotebox}

RQ$_2$ would allow us to understand if there is a need to study distinct settings of the context types or if any randomly chosen one can serve as a representative in the evaluation.

\subsection{Datasets}

\begin{table}
\caption{The real-world datatset used in fairness testing.}
\label{tb:dataset}
 \setlength{\tabcolsep}{0.7mm}
\begin{adjustbox}{width=\linewidth,center}
\begin{tabular}{l|l|l||l|l|l}
\toprule
    \textbf{Dataset} & \textbf{Domain} & \textbf{Size} & \textbf{Dataset} & \textbf{Domain} & \textbf{Size}\\
    \midrule
    \textsc{Adult ~\cite{adultdataset2017}} & Finance&$ 45,222$& \textsc{Student~\cite{studentperformancedataset2014}} & Education&$ 648$\\
   
    \textsc{Bank~\cite{bankdataset2014}} & Finance&$ 45,211$& \textsc{Crime~\cite{CommunitiesCrime2011}} & Criminology&$ 2,215$\\
   
    \textsc{German~\cite{germancreditdataset1994}} & Finance&$ 1,000$& \textsc{Kdd-Census~\cite{misc_census-income_(kdd)_117}} & Criminology&$ 284,556$\\
   
    \textsc{Credit~\cite{defaultcreditdataset2016}} & Finance&$30,000$&\textsc{Dutch~\cite{dutchcensus2014}} & Finance&$ 60,420$\\
   
    \textsc{Compas~\cite{compasdataset2016}} & Criminology&$ 6,172$&\textsc{Diabetes~\cite{diabetes1994}} & Healthcare&$ 45,715$\\
   
    \textsc{Law School~\cite{lawschooldataset1998}} & Education&$ 20,708$&\textsc{OULAD~\cite{oulad2017}} & Education&$ 21,562$\\

    \bottomrule
\end{tabular}
\end{adjustbox}
\end{table}


From the 53 papers surveyed in Section~\ref{sec:slr}, we followed the criteria below to select the tabular datasets for investigation:

\begin{itemize}
    \item For reproducibility, the datasets should be publicly available.
    \item To mitigate bias, the datasets come from diverse domains.
    \item The dataset should have been used by more than one peer-reviewed paper, and hence we rule out less reliable data.
    \item The datasets are free of missing values and repeated entries. 
\end{itemize}

As such, we selected 12 datasets as in Table~\ref{tb:dataset}.

\subsection{Metrics}

\subsubsection{Fairness Metric}

Since fairness testing for deep learning systems often works on individual fairness, we found that the percentage of Individual Discriminatory Instance\footnote{In this work, we use ``instance'' and ``sample'' interchangeably.} (IDI \%) among the generated samples is predominantly used as a way to measure to what extent the generated test cases can reveal fairness issues. In a nutshell, IDI assesses the extent to which two similar samples that only
differ in the sensitive attribute(s) but show different outcomes from the deep learning system, which exhibits individual discrimination---an equivalent measurement of the test oracle~\cite{DBLP:conf/issta/XiaoLL023,DBLP:conf/sigsoft/GalhotraBM17}. IDI \% would be calculated as ${\mathbb{I} / \mathbb{R}}$ where $\mathbb{I}$ is the number of IDI samples and $\mathbb{R}$ is the total number of sample created by a test generator.



\begin{table}[t!]
 \caption{Test adequacy metrics and commonalities (a paper might involve multiple metrics). The \setlength{\fboxsep}{1.5pt}\colorbox{teal!20}{green cells} indicate the chosen ones in our study.}
    \centering
    \setlength{\tabcolsep}{0.5mm}
    \begin{adjustbox}{width=\linewidth,center}
    \begin{tabular}{l|l|l|l}
    \toprule
          \textbf{Test Adequacy Metric} & \textbf{\# Papers} & \textbf{Optimality} & \textbf{Category}\\
          \midrule
          \rowcolor{teal!20}{Neuron Coverage (NC) ~\cite{pei2017deepxplore}} & 41 & Maximize & Coverage\\    
          \rowcolor{teal!20}{Strong Neuron Activation Coverage (SNAC)~\cite{ma2018deepgauge}}& 41 & Minimize & Coverage\\
          \rowcolor{teal!20}{Neuron Boundary Coverage (NBC)~\cite{zhou2020deepbillboard}} & 41 & Minimize & Coverage\\
          K-multisection Neuron Coverage (KNC)~\cite{yan2020correlations} & 38 & Maximize & Coverage \\
          Top-k Dominant Neuron Coverage (TKNC)~\cite{usman2022overview} & 30 & Maximize & Coverage\\
          Top-k Dominant Neuron Patterns Coverage (TKNP)~\cite{usman2022overview} & 30 & Maximize & Coverage\\
          \rowcolor{teal!20}{Distance-based Surprise Adequacy (DSA)~\cite{kim2019guiding}} & 15 & Maximize & Surprise\\
          \rowcolor{teal!20}{Likelihood-based Surprise Adequacy (LSA)~\cite{kim2019guiding}} & 15 & Maximize & Surprise\\   
          \bottomrule
    \end{tabular}
    \end{adjustbox}
   
    \label{tab:metric}
    
\end{table}

\subsubsection{Test Adequacy Metric}
\label{sec:tam}

Unlike the fairness metric, test adequacy metrics determine when the testing stops and it is highly common when testing certain properties of a deep learning system~\cite{DBLP:journals/tosem/ChenWMSSZC23,DBLP:journals/corr/abs-2207-10223}. Compared with directly using IDI to guide the testing, our preliminary results have revealed that the test adequacy metric can improve up to $\approx20$\% in discovering fairness bugs. The key reason is that the conventional adequacy metrics are highly quantifiable and provide more fine-grained, discriminative values for the test generation while fairness metrics like IDI are often more coarse-grained, similar to the case of using code coverage in classic software testing. This has been supported by prior work~\cite{DBLP:journals/corr/abs-2111-08856} and well-echoed by a well-known survey~\cite{DBLP:journals/corr/abs-2207-10223}.

In general, the adequacy metrics can be divided into \textit{coverage} metrics and \textit{surprise} metrics: the former refers to the coverage of the neuron activation as the test cases are fed into the DNN; the latter measures the diversity/novelty among the test cases. To understand which metrics are relevant, we analyzed the commonality of metrics from papers identified in Section~\ref{sec:slr}, as shown in Table~\ref{tab:metric}.



Since there are only two surprise metrics, we consider both DSA and LSA. Among the coverage metrics, some of them measure very similar aspects, e.g., NC, KBC, TKNC, and TKNP all focus on the overall neuron coverage while SNAC and NBC aim for different boundary areas/values of the neurons. As a result, in addition to SNAC and NBC, we use NC as the representative of KBC, TKNC, and TKNP since it is the most common metric among others.

Note that since test adequacy influences the resulting IDI, for each test generator, we examine its results on five test adequacy metrics and the independent IDI under each thereof.


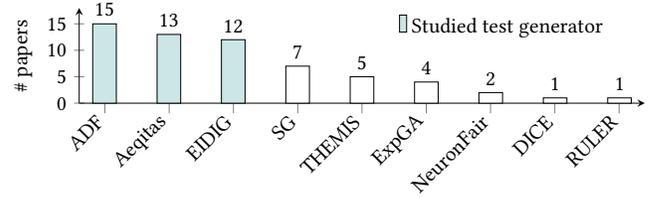
\begin{figure}
    \centering
    \begin{tikzpicture}
\begin{axis}[
    axis x line  = bottom,
    axis y line  = left  ,
    ybar=-0.35cm, 
    enlarge x limits=0.05,
    symbolic x coords={{ADF},{Aeqitas},{EIDIG},{SG},{THEMIS},{ExpGA},{NeuronFair},{DICE},{RULER}},
    xtick={{ADF},{Aeqitas},{EIDIG},{SG},{THEMIS},{ExpGA},{NeuronFair},{DICE},{RULER}},
    ylabel=\# papers,
    legend style={at={(0.75,1)},draw=none,anchor=north,legend columns=1},
    legend cell align={left},
    x tick label style={rotate=45,anchor=east},
    nodes near coords,
    height=3cm,
    width=10cm,
    ymax=18,
    ymin=0
    ]

\addplot [fill=teal!20,draw=black] coordinates {({ADF}, 15)};
\addplot [fill=teal!20,draw=black] coordinates {({Aeqitas}, 13)};
\addplot [fill=teal!20,draw=black] coordinates {({EIDIG}, 12)};

\addplot [fill=white,draw=black] coordinates {({SG}, 7)};
\addplot [fill=white,draw=black] coordinates {({THEMIS}, 5)};
\addplot [fill=white,draw=black] coordinates {({ExpGA}, 4)};
\addplot [fill=white,draw=black] coordinates {({NeuronFair}, 2)};
\addplot [fill=white,draw=black] coordinates {({DICE}, 1)};
\addplot [fill=white,draw=black] coordinates {({RULER}, 1)};


\addlegendentry{Studied test generator}
\end{axis}
\end{tikzpicture}
    \caption{The commonality of test generators evaluated in fairness testing (a paper might involve multiple generators). The y-axis shows the relevant paper count.}
    \label{fig:generator}
\end{figure}

\subsection{Test Generators}



To select generators for our study of fairness testing, we summarize the popularity of the generators from the 53 most relevant studies identified in Section~\ref{sec:slr}, as shown in Figure~\ref{fig:generator}. We use the top three most prevalent and representative generators, \texttt{Aequitas}~\cite{galhotra2017fairness}, \texttt{ADF}~\cite{zhang2020white}, and \texttt{EIDIG}~\cite{DBLP:conf/issta/ZhangZZ21}, which often serve as the state-of-the-art competitors in prior work. We consolidate those generators with the test adequacy metrics from Section~\ref{sec:tam}. All generators use a budget of $1,000$, which is sufficient for our study.

\begin{table}[]
    \caption{Context settings for hyperparameters. $[R]$, $[N]$, and $[C]$ denote real, integer, and categorical values, respectively.}
       \centering
       \setlength{\tabcolsep}{0.7mm}
    \begin{adjustbox}{width=\linewidth,center}
    \begin{tabular}{l|l||l|l}
             \toprule
        \textbf{Parameter}&\textbf{Range} &   \textbf{Parameter}&\textbf{Range}\\
     \midrule
           \text{learning\_rate$^{[R]}$}&$[10^{-6},10^{-1}]$&\text{optimizer$^{[C]}$}&$\{$Adam, SGD$\}$\\
           \text{batch\_size$^{[N]}$}&$\{2^4,2^5,...,2^{10}\}$&\text{num\_layers$^{[N]}$}&$[2,10]$\\
           \text{num\_epochs$^{[N]}$}&$[50,200]$&\text{regularization\_param$^{[R]}$}&$[10^{-5},10^{-1}]$\\
           \text{num\_units$^{[N]}$}&$\{2^4,2^5,...,2^{10}\}$&\text{weight\_init\_method$^{[C]}$}&$\{$Random, Xavier, HE$\}$\\
           \text{activation\_fn$^{[C]}$}&$\{$ReLU, Sigmoid, Tah$\}$&&\\

           \bottomrule
    \end{tabular}
        \end{adjustbox}

    \label{tab:hp-setting}
\end{table}

\subsection{Context Settings}

To mimic the possible real-world scenarios, we consider diverse settings for the three context types studied. As shown in Table~\ref{tab:hp-setting}, we study nine hyperparameters of DNN with the common value range used in the literature~\cite{DBLP:journals/corr/abs-1905-05786,DBLP:conf/kdd/Corbett-DaviesP17,DBLP:journals/tse/Nair0MSA20,DBLP:journals/corr/abs-2011-11001}. A configuration serves as a context setting thereof. For selection bias, we change the values of sensitive attributes such that the amount of data with a particular value and under the same label constitutes a random percentage, leading to a different context setting. Table~\ref{tab:sb-setting} shows the details and the ranges of percentages we can vary, which are aligned with the literature~\cite{DBLP:conf/fat/YeomT21,DBLP:conf/fairware-ws/ChakrabortyMT22,DBLP:conf/wacv/KarkkainenJ21}. This emulates different settings under which the extent of selection bias is varied. For example, in the \textsc{Diabetes} dataset~\cite{diabetes1994}, it is possible to provide two different settings\footnote{For datasets with multiple sensitive attributes, we consider all the possible combinations of the attributes' values.}:

\begin{itemize}

    \item One with $15\%$ samples as white on \textit{Race} under positive label while the remaining $85\%$ non-white \textit{Race} has negative label.
    \item Another with $30\%$ white \textit{Race} of negative label and the rest $70\%$ of non-white \textit{Race} come with positive label.
\end{itemize}

\begin{table}[]
    \caption{Context settings for selection bias.}
    \centering
    \begin{adjustbox}{width=\linewidth,center}
    \begin{tabular}{l|l|l}
    \toprule
          \textbf{Dataset}&\textbf{Sensitive Attribute(s) and Values} &\textbf{$\%$ Range}\\
    \midrule
    \multirow{3}{*}{\textsc{Adult}}& Sex (male and female)&$[10\%-90\%]$ \\
    &Race (white and non-white)&$[10\%-90\%]$\\
     &Age ($<25$, $25-65$, and $>65$)&$[10\%-80\%]$\\
    \hline
     \multirow{2}{*}{\textsc{Bank}}& Age ($<25$, $25-65$, and $>65$)&$[10\%-80\%]$\\
     &Marital (married and non-married)&$[10\%-90\%]$\\
    \hline
     \multirow{2}{*}{\textsc{German}}&Sex (male and female)&$[10\%-90\%]$ \\
     &Age ($<=25$ and $>25$)&$[10\%-90\%]$\\
    \hline
     \multirow{3}{*}{\textsc{Credit}}&Sex (male and female)&$[10\%-90\%]$ \\
     &Marital (married and non-married)&$[10\%-90\%]$\\
     &Education (university, high school, and others)&$[10\%-80\%]$\\
    \hline
     \multirow{2}{*}{\textsc{Compas}}&Sex (male and female)&$[10\%-90\%]$ \\
     &Race (white and non-white)&$[10\%-90\%]$\\
    \hline
     \multirow{2}{*}{\textsc{Law School}}&Sex (male and female)&$[10\%-90\%]$ \\
     &Race (white and non-white)&$[10\%-90\%]$\\
    \hline
     \multirow{2}{*}{\textsc{Student}}&Sex (male and female)&$[10\%-90\%]$ \\
     &Age ($<=18$ and $>18$)&$[10\%-90\%]$\\  
    \hline
     \multirow{2}{*}{\textsc{Crime}}&Race (white and non-white)&$[10\%-90\%]$\\
     & Age ($<=18$ and $>18$)&$[10\%-90\%]$\\
    \hline
     \multirow{2}{*}{\textsc{Kdd-Census}}&Sex (male and female)&$[10\%-90\%]$ \\
     &Race (white and non-white)&$[10\%-90\%]$\\
    \hline
     \multirow{1}{*}{\textsc{Dutch}}&Sex (male and female)&$[10\%-90\%]$ \\
    \hline
     \multirow{1}{*}{\textsc{Diabetes}}&Race (white and non-white)&$[10\%-90\%]$ \\
    \hline
     \multirow{1}{*}{\textsc{OULAD}}&Race (white and non-white)&$[10\%-90\%]$ \\

    \bottomrule
    \end{tabular}
    \end{adjustbox}
    \label{tab:sb-setting}
\end{table}

\begin{table}[t!]
    \caption{Context settings for label bias.}
    \centering
    \begin{adjustbox}{width=\linewidth,center}
    \begin{tabular}{l|l||l|l}
    \toprule
          \textbf{Dataset} &\textbf{Positive Class \% Range}& \textbf{Dataset} &\textbf{Positive Class \% Range}\\
         \midrule
    \textsc{Adult}&${[25\%-75\%]}$&\textsc{Student} &${[25\%-75\%]}$ \\

    \textsc{Bank} &${[12.5\%-87.5\%]}$& \textsc{Crime} &${[25\%-75\%]}$\\
    
    \textsc{German} &${[33\%-67\%]}$&\textsc{Kdd-Census} &${[5\%-95\%]}$\\

    \textsc{Credit} &${[25\%-75\%]}$&\textsc{Dutch} &${[25\%-75\%]}$\\

    \textsc{Compas} &${[33\%-67\%]}$&\textsc{Diabetes} &${[25\%-75\%]}$\\

    \textsc{Law School}&${[10\%-90\%]}$&\textsc{OULAD} &${[25\%-75\%]}$\\






    \bottomrule
    \end{tabular}
    \end{adjustbox}
    \label{tab:lb-setting}
\end{table}

Similarly, for label bias, we randomly alter the number of samples with positive/negative labels that are associated with the sensitive attributes using percentage bounds to create different context settings for it. Table~\ref{tab:lb-setting} shows the range of percentages as what has been used in existing work~\cite{DBLP:journals/corr/abs-2110-01109,DBLP:journals/corr/abs-2108-08504,compasdataset2016}. Taking the \textsc{Diabetes} dataset as an example again, two context settings can be one with a positive/negative ratio as $33\%:67\%$ while the other with $65\%:35\%$ for samples that have white and non-white values on \texttt{Race}.

\subsection{Protocol}

\begin{figure}
    \centering
    \includegraphics[width=\columnwidth]{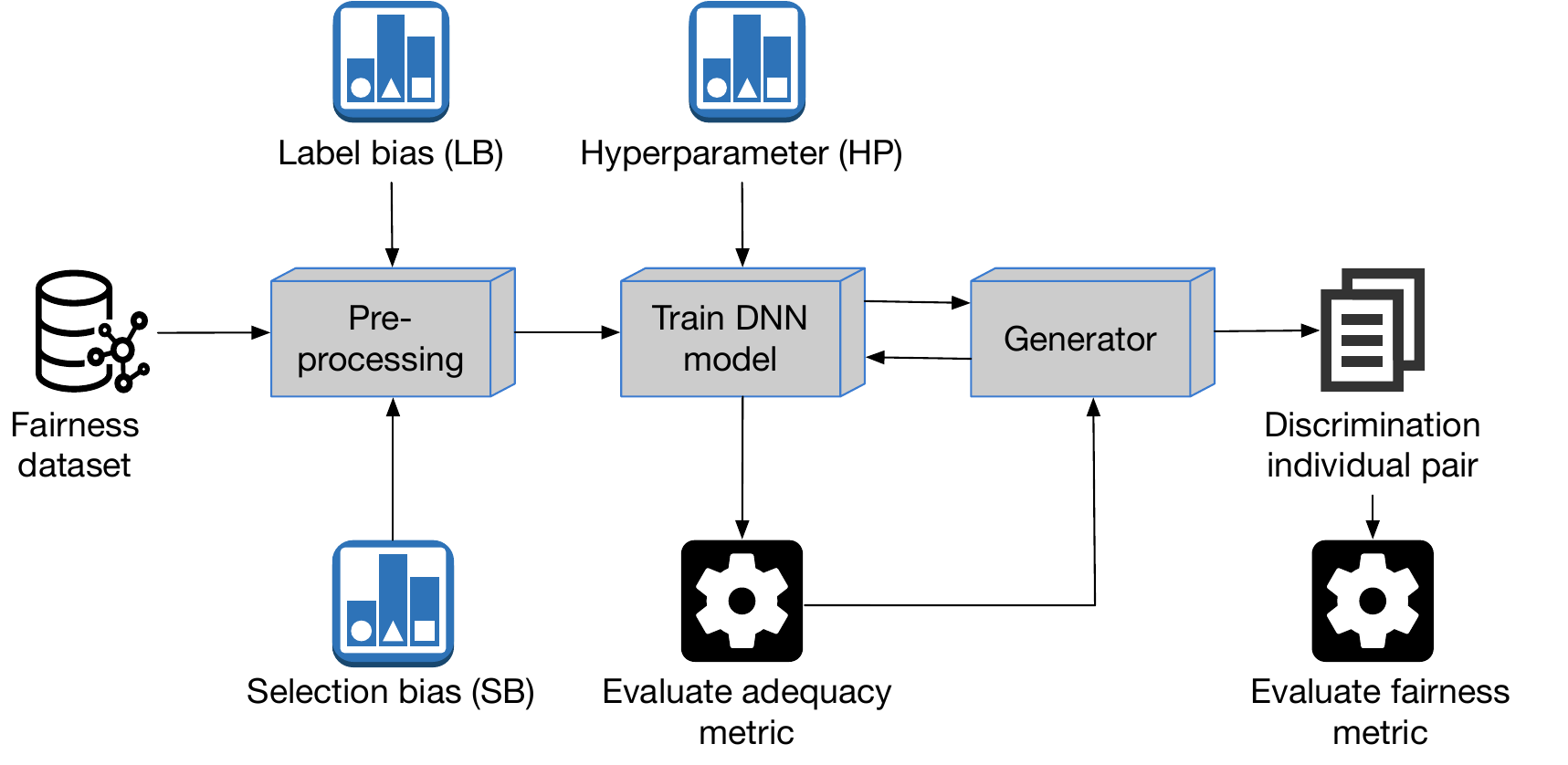}
    \caption{The protocol of our empirical study}
    \label{fig:protocol}
\end{figure}

Drawing on the above, we performed extensive experiments to assess how context changes can influence the fairness testing of a deep learning system, as shown in Figure~\ref{fig:protocol}. To avoid unnecessary noise, we build the three context types as follows:

\begin{itemize}
    \item \textbf{HP:} Only the hyperparameters settings are randomly changed; bias at the features and label is removed.
    \item \textbf{SB:} The selection bias is kept while the label bias is removed and the optimized hyperparameters are used~\cite{DBLP:journals/corr/abs-1905-05786,chakraborty2020making,gohar2023towards,tizpaz2022fairness}.
    \item \textbf{LB:} The label bias is varied but with removed selection bias and optimized hyperparameters.
    
\end{itemize}

For each context type, we use Latin Hypercube sampling~\cite{stein1987large} to sample 10 representative context settings for the analysis. In particular, SB and LB will affect the pre-processing, e.g., normalization and data cleaning, while HP runs with the model training. 

We compare the context types with a perfect/ideal \textit{baseline} that comes with optimized hyperparameters for accuracy while removing any selection and label bias, which is in line with the common assumptions in current work~\cite{DBLP:journals/corr/abs-1905-05786,chakraborty2020making,gohar2023towards,tizpaz2022fairness,chen2022maat,chakraborty2020fairway,chakraborty2021bias,DBLP:conf/wacv/KarkkainenJ21,mambreyan2022dataset,zhang2021ignorance,DBLP:conf/fairware-ws/ChakrabortyMT22,DBLP:journals/corr/abs-2108-08504}; this helps us to understand if considering changes in a context type would alter the results compared with the case when none of the contexts is considered. It is worth noting that it took a huge effort to completely remove the selection and label bias~\cite{DBLP:conf/iccv/ChenJ21,DBLP:conf/wacv/KarkkainenJ21}. Similarly, tuning the configuration of hyperparameters is also known to be expensive~\cite{DBLP:conf/kbse/LiXCT20,DBLP:journals/tosem/ChenL23a,DBLP:journals/tosem/ChenL23,feurer2019hyperparameter,DBLP:journals/tse/ChenCL24,DBLP:journals/pacmse/0001L24,DBLP:conf/sigsoft/0001L21,DBLP:conf/wcre/Chen22,DBLP:journals/tosem/ChenLBY18}, even though some surrogates might be used~\cite{DBLP:conf/sigsoft/Gong023,DBLP:journals/pacmse/Gong024}. Hence, we set the ideal situations only for the purpose of our study and they might not be desirable in practice.

\begin{table*}[t!]
\caption{The average Scott-Knott rank scores for the quality metric (5 cases); the lower the rank score, the better. \setlength{\fboxsep}{1.5pt}\colorbox{teal!20}{green cells} and \setlength{\fboxsep}{1.5pt}\colorbox{red!20}{red cells} means the rank is better and worse than the baseline in a dataset-generator pair, respectively.}

\label{tb:rq1}
\setlength{\tabcolsep}{0.7mm}
\centering
\begin{adjustbox}{width=\linewidth,center}
\begin{tabular}{l|cccc|cccc|cccc||cccc|cccc|cccc}\toprule

\multirow{3}{*}{\textbf{Dataset}}&\multicolumn{12}{c||}{\textbf{Test Adequacy Metrics}}&\multicolumn{12}{c}{\textbf{Fairness Metric}}\\

\cmidrule{2-25}

&\multicolumn{4}{c|}{\textbf{\texttt{Aequitas} }}&\multicolumn{4}{c|}{\textbf{\texttt{ADF}}}&\multicolumn{4}{c||}{\textbf{\texttt{EIDIG}}}&\multicolumn{4}{c|}{\textbf{\texttt{Aequitas}}}&\multicolumn{4}{c|}{\textbf{\texttt{ADF}}}&\multicolumn{4}{c}{\textbf{\texttt{EIDIG}}}\\

\cmidrule{2-25}

&Baseline&HP&SB&LB&Baseline&HP&SB&LB&Baseline&HP&SB&LB&Baseline&HP&SB&LB&Baseline&HP&SB&LB&Baseline&HP&SB&LB\\

\midrule

\textsc{Bank}&3.2&\colorbox{red!20}{3.8}&\colorbox{teal!20}{2.0}&\colorbox{teal!20}{1.8}&2.6&\colorbox{red!20}{3.8}& \colorbox{teal!20}{1.8}&\colorbox{teal!20}{1.8}&2.6&\colorbox{red!20}{3.6}&\colorbox{teal!20}{1.8} &\colorbox{teal!20}{1.8}&2.4 &\colorbox{red!20}{2.8} &{\colorbox{teal!20}{2.0}}&\colorbox{red!20}{3.4}&2.4 &\colorbox{red!20}{2.8}&\colorbox{teal!20}{1.8} &\colorbox{red!20}{3.2}&2.0 &\colorbox{red!20}{2.4} &\colorbox{teal!20}{1.8}&\colorbox{red!20}{3.0} \\

\textsc{Adult}& 3.2& \colorbox{red!20}{3.8}& \colorbox{teal!20}{2.6}& \colorbox{teal!20}{1.8}& 2.8& \colorbox{red!20}{3.4}& \colorbox{teal!20}{2.2}& \colorbox{teal!20}{1.6} & 2.6& \colorbox{red!20}{3.2}& \colorbox{teal!20}{2.0} & \colorbox{teal!20}{1.6}& 2.8 & \colorbox{teal!20}{2.4}& \colorbox{teal!20}{2.2} & \colorbox{red!20}{2.8} & 2.6 & \colorbox{teal!20}{2.4}& \colorbox{teal!20}{2.0}& \colorbox{red!20}{2.8}  & 2.6 & \colorbox{teal!20}{2.2} & \colorbox{teal!20}{2.0}& \colorbox{teal!20}{2.4} \\

\textsc{Credit}& 3.2 & \colorbox{teal!20}{2.6}& \colorbox{teal!20}{2.8} & \colorbox{teal!20}{3.0}  & 3.0 & \colorbox{teal!20}{2.2} & \colorbox{teal!20}{2.2} & \colorbox{teal!20}{2.2}& 1.6 & \colorbox{red!20}{2.8}  & \colorbox{red!20}{2.2} & \colorbox{red!20}{2.2}&2.8 &\colorbox{red!20}{3.2} & \colorbox{teal!20}{1.6}& 2.8  & 2.4 & \colorbox{red!20}{3.0}& \colorbox{teal!20}{1.6} & \colorbox{red!20}{2.6}  & 2.2 & \colorbox{red!20}{2.6}& \colorbox{teal!20}{1.4} & \colorbox{red!20}{2.4} \\

\textsc{German}& 3.4 & \colorbox{red!20}{3.6}& \colorbox{teal!20}{2.0} & \colorbox{teal!20}{2.2}  & 3.2 & \colorbox{teal!20}{2.8} & \colorbox{teal!20}{2.2}& \colorbox{teal!20}{2.6} & 2.8 & \colorbox{teal!20}{2.6}  & \colorbox{teal!20}{2.0}& \colorbox{teal!20}{2.2} & 2.8 & \colorbox{teal!20}{2.2}  & \colorbox{teal!20}{2.6}& \colorbox{teal!20}{2.2} & 2.6 & \colorbox{teal!20}{2.2} & 2.6 & \colorbox{teal!20}{2.2} & 2.2 & 2.2& \colorbox{teal!20}{2.0} & \colorbox{red!20}{2.8} \\

\textsc{Compas}& 3.2 & \colorbox{red!20}{3.4} & \colorbox{teal!20}{1.8}& \colorbox{teal!20}{2.4}  & 2.8 & \colorbox{red!20}{3.0} & \colorbox{teal!20}{1.6}& \colorbox{teal!20}{2.0}  & 2.4 & \colorbox{red!20}{3.0}  & \colorbox{teal!20}{1.4}& \colorbox{teal!20}{2.0}& 2.6 & \colorbox{teal!20}{2.2}  & \colorbox{red!20}{3.2}& \colorbox{teal!20}{2.2} & 2.6 & \colorbox{teal!20}{2.2} & \colorbox{red!20}{3.2}& \colorbox{teal!20}{2.2} & 2.4 & \colorbox{teal!20}{2.0} & \colorbox{red!20}{2.6}& \colorbox{teal!20}{2.0} \\

\textsc{Law School}& 3.2 & \colorbox{red!20}{3.2} & \colorbox{teal!20}{3.0} & \colorbox{teal!20}{1.8} & 3.2 &\colorbox{red!20}{3.6} & \colorbox{teal!20}{2.4}& \colorbox{teal!20}{1.4} & 2.8 & \colorbox{red!20}{3.4} & \colorbox{teal!20}{2.4}& \colorbox{teal!20}{1.4}& 3.2 & \colorbox{teal!20}{2.6}& \colorbox{teal!20}{2.4} & \colorbox{teal!20}{1.8} & 3.0 & \colorbox{teal!20}{2.6} & \colorbox{teal!20}{2.4} & \colorbox{teal!20}{1.6} & 2.8 & \colorbox{teal!20}{2.6} & \colorbox{teal!20}{1.8}& \colorbox{teal!20}{2.2}\\

\textsc{Student}& 3.0 & \colorbox{teal!20}{2.2} & \colorbox{red!20}{3.2} & \colorbox{teal!20}{2.6} & 2.6 & \colorbox{teal!20}{2.0} & \colorbox{red!20}{3.2} & 2.6 & 2.2 & \colorbox{teal!20}{2.0} & \colorbox{red!20}{3.2}& \colorbox{red!20}{2.4} &1.8 & \colorbox{red!20}{2.6}& \colorbox{red!20}{2.8}  & \colorbox{teal!20}{2.4} & 1.8 & \colorbox{red!20}{2.6} & \colorbox{red!20}{2.8}& \colorbox{teal!20}{2.4}  & 1.6 & \colorbox{red!20}{2.6}& \colorbox{red!20}{2.6}& \colorbox{teal!20}{2.4}\\

\textsc{Crime}& 3.0 & \colorbox{red!20}{3.8} & \colorbox{teal!20}{2.0} & \colorbox{teal!20}{1.6} & 3.0 & \colorbox{red!20}{3.6}  & \colorbox{teal!20}{2.0}& \colorbox{teal!20}{1.4} & 2.6 & \colorbox{red!20}{3.2}  & \colorbox{teal!20}{2.0}& \colorbox{teal!20}{1.4}& 2.8 & \colorbox{red!20}{3.0} & \colorbox{teal!20}{2.4} & \colorbox{teal!20}{2.2}& 2.2 & \colorbox{red!20}{2.4} & \colorbox{teal!20}{2.0}& \colorbox{teal!20}{1.8} & 2.0 & \colorbox{red!20}{2.2} & 2.0& \colorbox{teal!20}{1.8}\\

\textsc{Kdd-Census}& 2.8 & \colorbox{red!20}{2.8} & \colorbox{red!20}{2.8}& \colorbox{teal!20}{2.2}  & 2.2 & \colorbox{red!20}{2.8}& \colorbox{red!20}{3.0} & \colorbox{teal!20}{2.4}  & 2.0 & \colorbox{red!20}{2.4} & \colorbox{red!20}{2.6}& \colorbox{teal!20}{2.0}&3.0 & \colorbox{teal!20}{2.6}  & \colorbox{teal!20}{2.8}& \colorbox{teal!20}{2.0} & 2.6 & 2.6  & 2.6 & \colorbox{teal!20}{2.0}& 2.6 & 2.6 & 2.6 & \colorbox{teal!20}{2.0} \\

\textsc{Dutch}& 2.8 & \colorbox{red!20}{4.4}  & \colorbox{teal!20}{2.2} & \colorbox{teal!20}{2.2}& 2.8 & \colorbox{red!20}{3.4}  & \colorbox{teal!20}{1.6}& \colorbox{teal!20}{2.0} & 2.6 & \colorbox{red!20}{3.2}  & \colorbox{teal!20}{1.6}& \colorbox{teal!20}{2.0}& 2.4 & \colorbox{teal!20}{2.2}  & \colorbox{red!20}{3.2}& 2.4 & 2.2 & 2.2 & \colorbox{red!20}{3.2} & \colorbox{red!20}{2.6}& 2.0 & \colorbox{red!20}{2.2} & \colorbox{red!20}{2.6}& \colorbox{red!20}{2.4} \\

\textsc{Diabetes}& 2.6 & \colorbox{red!20}{3.6}  & \colorbox{teal!20}{2.0} & \colorbox{teal!20}{2.2}& 3.2 & \colorbox{red!20}{3.2}  & \colorbox{teal!20}{2.0} & \colorbox{teal!20}{2.0}& 2.8 & \colorbox{red!20}{3.0}  & \colorbox{teal!20}{1.8} & \colorbox{teal!20}{2.0}&2.6 & \colorbox{red!20}{3.0}  & \colorbox{teal!20}{2.0}& \colorbox{teal!20}{2.2} & 2.4 & \colorbox{red!20}{3.0}  & \colorbox{teal!20}{2.0}& \colorbox{teal!20}{2.2} & 2.0 & \colorbox{red!20}{2.8} & 2.0 & \colorbox{red!20}{2.2}\\

\textsc{OULAD}& 3.8 &\colorbox{red!20}{4.0} & \colorbox{teal!20}{2.6}& \colorbox{teal!20}{2.0}  & 3.0 & \colorbox{red!20}{3.4}  & \colorbox{teal!20}{2.0} & \colorbox{teal!20}{1.4} & 2.4 & \colorbox{red!20}{3.4}  & \colorbox{teal!20}{1.8}& \colorbox{teal!20}{1.4}&3.0 & \colorbox{teal!20}{2.8} & \colorbox{teal!20}{1.6}&\colorbox{teal!20}{2.8} & 2.8 & \colorbox{teal!20}{2.6}  & \colorbox{teal!20}{1.6} & \colorbox{teal!20}{2.4}& 2.2 & \colorbox{red!20}{2.6} & \colorbox{teal!20}{1.4}& \colorbox{teal!20}{2.0} \\

\bottomrule
\end{tabular}
\end{adjustbox}
    \vspace{-0.3cm}
\end{table*}

We run all three context types with 12 datasets, three generators, 10 context settings, and five test adequacy metrics together with IDI as the fairness metric. The training/testing data split follows $70\%$/$30\%$~\cite{DBLP:conf/icse/LiX0WT20}. We repeat each experiment for 30 runs.

\section{Results}
\label{sec:results}

In this section, we illustrate and discuss the results of our study.



\subsection{Contextual Implication to Metrics (RQ$_1$)}

\subsubsection{Method}


For RQ$_1$, we compare baseline, HP, SB, and LB paired with each generator, leading to 12 subjects. To ensure statistical significance, for each dataset, we report on the mean rank scores from the Scott-Knott test\footnote{Scott-Knott test~\cite{DBLP:journals/tse/MittasA13} is a statistical method that ranks the subjects.}~\cite{DBLP:journals/tse/MittasA13} on 12 subjects over the five metrics. The 10 context settings and 30 repeated runs lead to $10\times30=300$ data points in Scott-Knott test per subject, in which the data from the baseline is repeated as it is insensitive to context settings.  





\subsubsection{Results}

For test adequacy metrics, as shown in Table~\ref{tb:rq1}, we see that testing with biased selection and label of data have better ranks in general compared with their baseline, leading to 29 (81\%) and 33 (92\%) better cases out of the 36 dataset-generator pairs, respectively. Under non-optimized hyperparameters, in contrast, there are 29 (81\%) cases worse than the baseline. This means that generators would struggle to reach better adequacy results under non-optimized hyperparameters while the presence of selection/label bias would boost their performance. 



From Table~\ref{tb:rq1}, we also compare the results on the fairness metric. Clearly, across all cases, testing under non-optimized hyperparameters leads to merely 15/36 (42\%) cases that exhibit more effective outcomes in finding fairness bugs than the baseline. On the contrary, this number becomes 22 (61\%) and 23 (64\%) for testing under selection and label bias, respectively, which is again rather close. 

Interestingly, the contexts tend to change the comparative results between generators. For example, on the \textsc{Diabetes} dataset with test adequacy metrics, the rank score under baseline for \texttt{Aequitas}, \texttt{ADF}, and \texttt{EIDIG} is 2.6, 3.2, and 2.8, respectively, hence \texttt{Aequitas} tends to be the best. However, HP changes it to 3.6, 3.2, and 3.0 (hence \texttt{ADF} performs better than \texttt{Aequitas}) while LB makes it as 2.2, 2.0 and 2.0 (hence both \texttt{ADF} and \texttt{EIDIG} perform the best). This suggests that the contexts, if explicitly considered, can invalidate the conclusions drawn in existing work under the baseline setting.

Another pattern we found is that, for all context types, there is a moderate decrease in the performance deviation to baseline on fairness bug discovery when compared with the cases for test adequacy metrics, but the overall conclusion remains similar. This implies that a better adequacy metric value might not always lead to a better ability to find fairness bugs across contexts. 




   


\begin{quotebox}
   \noindent
   \textit{\textbf{Response to RQ$_1$:} Compared with the ideal baseline where no context has flaws, non-optimized hyperparameters often cause the generators more struggle to improve test adequacy (81\% cases) and find fairness bugs (58\% cases); while the data bias generally allows current generators to perform better for all metrics (61\%-92\% cases). Contexts may also change the rankings between generators. }
\end{quotebox}

\subsection{Significance of Context Settings (RQ$_2$)}

\subsubsection{Method}

To understand RQ$_2$, in each context/dataset, we verify the statistically significant difference between the 10 context settings over all combinations of the generators and metrics. To avoid the lower statistical power caused by multiple comparison tests~\cite{wilson2019harmonic} (e.g., Kruskal-Wallis test~\cite{mckight2010kruskal} with Bonferroni correction~\cite{napierala2012bonferroni}) and the restriction of independent assumption (e.g., in Fisher's test~\cite{upton1992fisher}),  we at first perform pair-wise comparisons using Wilcoxon sum-rank test~\cite{mann1947test}---a non-parametric and non-paired test. We do so for all 45 pairs in the metric results of 10 context settings with 30 runs each. Next, we carry out correction using Harmonic mean $p$-value~\cite{wilson2019harmonic}, denoted as $p\!\:\!\!\circo$, since it does not rely on the independence of the $p$-values while keeping sufficient statistical power~\cite{wilson2019harmonic}. When the Harmonic mean $p\!\:\!\!\circo < 0.05$, we reject the hypothesis that the 45 pairs of metric results across the 10 context settings in a case are of no statistical difference~\cite{wilson2019harmonic,wilson2019harmonic1}. Since there are 12 datasets, three generators, and five metrics (for both adequacy and fairness), we have $12 \times 3 \times 5= 180$ cases in total.



\subsubsection{Result}

\begin{figure}[t]
     \centering
          \includegraphics[height=\columnwidth]{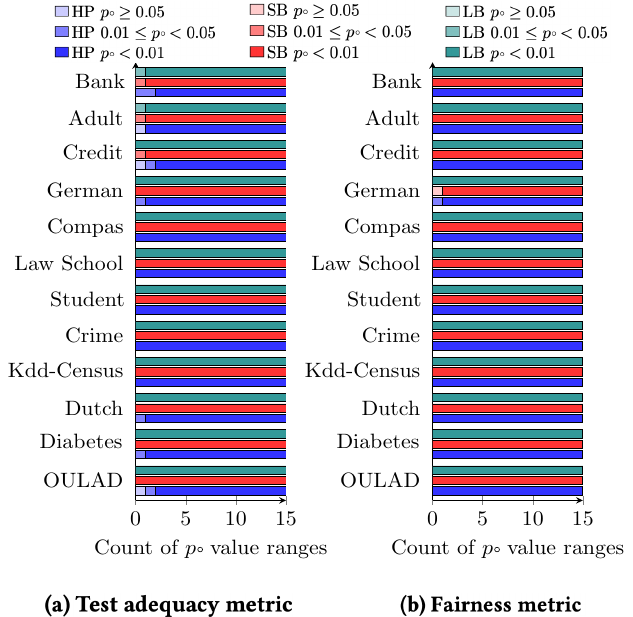}         
     \caption{The $\#$ cases with different ranges for the Harmonic mean $p$-value from the 15 cases of generators and metrics.}
    \label{fig:rq2}
\end{figure}

The results are plotted in Figure~\ref{fig:rq2} where we count the number of cases with different ranges of $p\!\:\!\!\circo$ within the 15 generator-metric combinations for each dataset. 

As can be seen, for the adequacy metrics, the majority of the cases are having $p\!\:\!\!\circo < 0.05$, most of which exhibit $p\!\:\!\!\circo < 10^{-7}$. We observe only a small fraction of cases with $p\!\:\!\!\circo \geq 0.01$, i.e., $13/180=7\%$, including 3 cases of $p\!\:\!\!\circo \geq 0.05$. This suggests that varying context settings have significant implications for the testing. For the fairness metric, we see even stronger evidence for the importance of context setting: there are only two cases with $0.01 \leq p\!\:\!\!\circo < 0.05$, and none of them have $p\!\:\!\!\circo \geq 0.05$. Again, $p\!\:\!\!\circo < 10^{-7}$ is a very common result.

Notably, we can also observe a discrepancy between the results of the adequacy metric and that of the fairness metric. This suggests some complex and non-linear relationships between the two metric types, which we will further explore in Section~\ref{sec:exploration}.


\begin{quotebox}
   \noindent
   \textit{\textbf{Response to RQ$_2$:} Within each context type, changing settings generally leads to significant impacts on the testing results in terms of both adequacy and fairness metrics.}
\end{quotebox}

\section{Causes of Context Sensitivity}
\label{sec:exploration}

We have revealed some interesting findings on the implication of contexts and their settings for fairness testing of deep learning systems. Naturally, the next question to study is: what are the key reasons behind the observed phenomena? To explore this, we ask:

\begin{quotebox}
   \noindent
   \textit{\textbf{RQ$_3$:} What problem characteristics of fairness testing have been affected under varying contexts and their settings?}
\end{quotebox}



A consistent pattern we found for RQ$_1$ and RQ$_2$ is that a better value on the test adequacy metric might not necessarily lead to a better fairness metric during testing. In fact, we observed a considerable discrepancy between them, particularly when non-optimized hyperparameters exist. Therefore, we also investigate:

\begin{quotebox}
   \noindent
   \textit{\textbf{RQ$_4$:} What is the correlation between a test adequacy metric and the fairness metric with respect to the contexts?}
\end{quotebox}

The answer to the above can serve as a foundation for vast future research directions for fairness testing of deep learning systems.

\subsection{Testing Landscape Analysis (RQ$_3$)}

\subsubsection{Method}


We leverage fitness landscape analysis~\cite{DBLP:series/sci/PitzerA12} to understand RQ$_3$ through well-established metrics that interpret the structure and difficulty of the landscape. Since the test generators explore the space of the test adequacy metrics, we focus on test adequacy in the fitness landscape. This is analogous to the classic testing where fitness analysis is also conducted in the code coverage space~\cite{DBLP:journals/tse/NeelofarSMA23}. Their correlations to the IDI are then studied in RQ$_4$.


Specifically, at the local level, we use correlation length ($\ell$)~\cite{stadler1996landscapes}---a metric to measure the local ruggedness of the landscape surface that a generator is likely to visit. Formally, $\ell$ is defined as:
\begin{align}
		\ell(p,s) = - (\ln|{1 \over {\sigma_f^2 (p-s)}} \sum^{p-s}_{i=1} (f_i - \overline{f}) (f_{i+s} - \overline{f})|)^{-1} 
		 \label{Eq:rugg}
\end{align}
$\ell$ is essentially a normalized autocorrelation function of neighboring points’ adequacy values explored; $f_i$ is the test adequacy of the $i$th test case visited by the random walks~\cite{DBLP:journals/tsmc/TavaresPC08}. $s$ denotes the step size and $p$ is the walk length ($p=100$). We use $s$ = 1 in this work, which is the most restricted neighborhood deinfition~\cite{DBLP:journals/tsmc/TavaresPC08,DBLP:conf/gecco/OchoaQB09}. The higher the value of $\ell$, the smoother the landscape, as the test adequacy of adjacently sampled test cases are more correlated~\cite{stadler1996landscapes}. 





At the global level, we use Fitness Distance Correlation $\varrho$ (FDC)~\cite{DBLP:conf/icga/JonesF95} to quantify the guidance of the landscape for the test generator:
\begin{align}
	 \varrho(f,d) = {1 \over {\sigma_f \sigma_d p}} \sum^p_{i=1} (f_i - \overline{f}) (d_i - \overline{d})
	 \label{Eq:fdc}
\end{align}
where $p$ is the number of points considered in FDC; in this work, we adopt Latin Hypercube sampling to collect 100 data points from the landscape. Within such a sample set, $d_i$ is the shortest Hamming distance of the $i$th test case to a global optimum we found throughout all experiments. $\overline{f}$ ($\overline{d}$)
and $\sigma_f$ ($\sigma_d$) are the mean and standard deviation of test adequacy (and distance), respectively. For maximized test adequacy metrics, $-1 \leq \varrho < 0$ implies that the distance to the global optimal decreases as the adequacy results become better, hence the landscape has good guidance for a generator. Otherwise, the landscape tends to be misleading. A higher $|\varrho|$ indicates a stronger correlation/guidance. 








Here, we treat each context setting under a context type as an independent testing landscape. This, together with 12 datasets and 5 test adequacy metrics, gives us $12\times5\times10=600$ different landscapes to analyze. To better study the variability of the landscape caused by different context settings, we also report the coefficient of variation (CV) for the landscape metrics over each case of 10 context settings, i.e. ${{\sigma_i} \over {\mu_i}}\times 100\%$, where $\mu_i$ and $\sigma_i$ are the mean and standard deviation on the FDC/correlation length values for 10 context settings in the $i$th case, respectively.  As such, we have $60$ CV values to examine.

\subsubsection{Result}

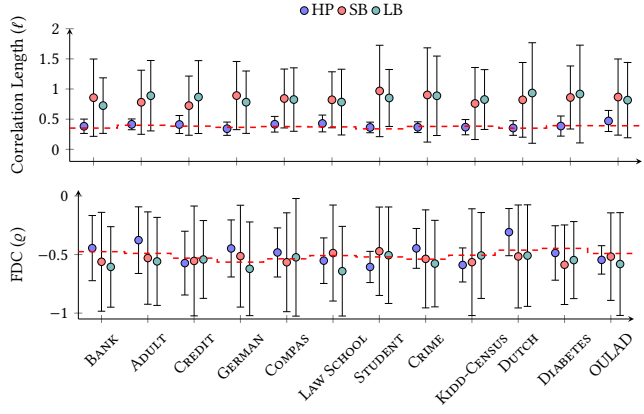
\begin{figure}[t!]
     \centering
     \begin{subfigure}[t]{\columnwidth}
          \centering
          \begin{tikzpicture}
\pgfplotsset{
  error bars/.cd,
    x dir=none,
    y dir=both, y explicit,
}
\begin{axis}[
    width=12cm,
    height=4cm,
    only marks,
      axis x line  = bottom,
    axis y line  = left  ,
    xmin=-0.5,xmax=11.5,ymin=-0.2,ymax=2,
    xtick={0,1,2,3,4,5,6,7,8,9,10,11},
    xticklabels={},
    x tick label style={rotate=45},
    ylabel=Correlation Length ($\ell$),
      legend columns=3,
     legend style={nodes={scale=1, transform shape},draw=none,fill=none,at={(0.6,1.25)}}
  ]
  \addplot+ [mark=*,mark options={fill=blue!50,mark size=2pt,draw=black},color=black] table [x expr={\thisrow{x}-0.2}, y=y, y error=ey] {
    x y ey
0	0.382357394	0.118919315
1	0.414215195	0.08934569
2	0.411761679	0.149322096
3	0.3419269	0.110215069
4	0.415837864	0.129964843
5	0.428635762	0.139307349
6	0.36298725	0.087988518
7	0.366910013	0.088914956
8	0.367272849	0.126286294
9	0.353306767	0.121149018
10	0.385932643	0.166668432
11	0.470201169	0.174612741
    
  };
  \addplot+ [mark=*,mark options={fill=red!50,mark size=2pt,draw=black},color=black] table [x expr={\thisrow{x}}, y=y, y error=ey] {
    x y ey
0	0.856233844	0.640791809
1	0.780264549	0.531176261
2	0.723973761	0.490465709
3	0.891249005	0.56515825
4	0.842726917	0.488233278
5	0.821128002	0.464256652
6	0.967575164	0.757631407
7	0.901399494	0.781021019
8	0.759678105	0.597078357
9	0.820327601	0.619138676
10	0.859467016	0.522475613
11	0.866627227	0.631035347   
  };
    \addplot+ [mark=*,mark options={fill=teal!50,mark size=2pt,draw=black},color=black] table [x expr={\thisrow{x}+0.2}, y=y, y error=ey] {
    x y ey
0	0.724628113	0.460414322
1	0.889182852	0.583683584
2	0.86652389	0.603739488
3	0.781431324	0.516987845
4	0.825187937	0.525813332
5	0.783037326	0.544921963
6	0.851595511	0.472315733
7	0.8875691	0.658425493
8	0.825592097	0.49573272
9	0.933688411	0.833403872
10	0.916759902	0.810268013
11	0.815750534	0.62416019
  };

    \legend{
HP,
SB,
LB
}
\end{axis}

\begin{axis}[hide axis, width=12cm,  every axis plot/.append style={thick},height=4cm,   xmin=-0.5,xmax=11.5,ymin=-0.2,ymax=2]
  \addplot [dashed,red] plot coordinates {
    (-0.5,0.351990309322818)
    (0.5,0.351990309322818)
    (0.5,0.398851608424099)
    (1.5,0.398851608424099)  
    (1.5,0.384277866216396)
    (2.5,0.384277866216396) 
    (2.5,0.364545396353382)
    (3.5,0.364545396353382) 
    (3.5,0.377433688764136)
    (4.5,0.377433688764136) 
    (4.5,0.371821936571434)
    (5.5,0.371821936571434) 
    (5.5,0.339564104780526)
    (6.5,0.339564104780526) 
    (6.5,0.37952231295787)
    (7.5,0.37952231295787) 
    (7.5,0.383805746592712)
    (8.5,0.383805746592712) 
    (8.5,0.350576952048015)
    (9.5,0.350576952048015) 
    (9.5,0.392618907414289)
    (10.5,0.392618907414289)
    (10.5,0.391029975394994)
    (11.5,0.391029975394994) 
};
  
\end{axis}

\end{tikzpicture}
     \end{subfigure}
     \vspace{-0.2cm}
     \begin{subfigure}[t]{\columnwidth}
          \centering
          \begin{tikzpicture}
\pgfplotsset{
  error bars/.cd,
    x dir=none,
    y dir=both, y explicit,
}
\begin{axis}[
    width=12cm,
    height=4cm,
    only marks,
      axis x line  = bottom,
    axis y line  = left  ,
    xmin=-0.5,xmax=11.5,ymin=-1.05,ymax=0.05,
    xtick={0,1,2,3,4,5,6,7,8,9,10,11},
    xticklabels={\textsc{Bank},\textsc{Adult},\textsc{Credit},\textsc{German},\textsc{Compas},\textsc{Law School},\textsc{Student},\textsc{Crime},\textsc{Kidd-Census},\textsc{Dutch},\textsc{Diabetes},\textsc{OULAD}},
    x tick label style={rotate=45},
    ylabel=FDC ($\varrho$),
    legend columns=3,
     legend style={nodes={scale=1, transform shape},draw=none,fill=none,at={(0.6,1.25)}}
  ]
  \addplot+ [mark=*,mark options={fill=blue!50,mark size=2pt,draw=black},color=black] table [x expr={\thisrow{x}-0.2}, y=y, y error=ey] {
    x y ey
  0	-0.445074811	0.278160807
1	-0.377172263	0.284404974
2	-0.573145799	0.272113082
3	-0.448025409	0.243423557
4	-0.482133062	0.209483304
5	-0.552407968	0.194469995
6	-0.605964839	0.133133915
7	-0.447419287	0.169974136
8	-0.588823731	0.145313486
9	-0.308649047	0.201254236
10	-0.486603812	0.232354529
11	-0.545709116	0.121242794
    
  };
  \addplot+ [mark=*,mark options={fill=red!50,mark size=2pt,draw=black},color=black] table [x expr={\thisrow{x}}, y=y, y error=ey] {
    x y ey
 0	-0.561750842	0.421887731
1	-0.53010101	0.394150658
2	-0.554612795	0.4687638
3	-0.513804714	0.434682612
4	-0.565791246	0.421918685
5	-0.487138047	0.409706865
6	-0.471919192	0.377188158
7	-0.536969697	0.418353245
8	-0.565252525	0.455599048
9	-0.516363636	0.43939059
10	-0.586936027	0.339499102
11	-0.517710438	0.37359512    
  };
    \addplot+ [mark=*,mark options={fill=teal!50,mark size=2pt,draw=black},color=black] table [x expr={\thisrow{x}+0.2}, y=y, y error=ey] {
    x y ey
0	-0.605656566	0.343618911
1	-0.558518519	0.37566563
2	-0.541952862	0.3319947
3	-0.621683502	0.399708404
4	-0.523771044	0.50180197
5	-0.642154882	0.382626144
6	-0.505723906	0.411792633
7	-0.577239057	0.369261984
8	-0.508148148	0.366601479
9	-0.508956229	0.433432794
10	-0.547474747	0.328781629
11	-0.581010101	0.438712223 
  };

\end{axis}
\begin{axis}[hide axis, width=12cm,  every axis plot/.append style={thick},height=4cm,xmin=-0.5,xmax=11.5,ymin=-1.05,ymax=0.05]
  \addplot [dashed,red] plot coordinates {
    (-0.5,-0.475400267299457)
    (0.5,-0.475400267299457)
    (0.5,-0.490145176133775)
    (1.5,-0.490145176133775)  
    (1.5,-0.530860358763149)
    (2.5,-0.530860358763149) 
    (2.5,-0.564715057364322)
    (3.5,-0.564715057364322) 
    (3.5,-0.536745391710888)
    (4.5,-0.536745391710888) 
    (4.5,-0.508824215754909)
    (5.5,-0.508824215754909) 
    (5.5,-0.520414061608181)
    (6.5,-0.520414061608181) 
    (6.5,-0.539388888383788)
    (7.5,-0.539388888383788) 
    (7.5,-0.505803307603488)
    (8.5,-0.505803307603488) 
    (8.5,-0.462434122200099)
    (9.5,-0.462434122200099) 
    (9.5,-0.447512225970072)
    (10.5,-0.447512225970072)
    (10.5,-0.491002635617097)
    (11.5,-0.491002635617097) 
};
\end{axis}


\end{tikzpicture}
     \end{subfigure}
     \vspace{-0.2cm}
     \caption{Mean and deviation of the FDC and correlation length over 50 cases of adequacy metrics and context settings per dataset. The dashed line is the value of the baseline.}
    \label{fig:rq4-1}
    \vspace{-0.5cm}
\end{figure}

In Figure~\ref{fig:rq4-1}, for correlation length, we see that the presence of data bias generally leads to a smoother landscape to be tested than the baseline; while testing on the non-optimized hyperparameters shares similar ruggedness to that of the baseline. When comparing with the baseline on FDC, the non-optimized hyperparameters often lead to a landscape with weaker guidance while selection and label bias generally show stronger guidance.

The above explains the observations from RQ$_1$: the context related to hyperparameters makes the testing problem at the model level more challenging due to the loss of fitness guidance while keeping the ruggedness similar. In contrast, the selection and label bias render stronger fitness guidance with a smoother landscape, leading to an easier-to-be-tested testing problem. 

\begin{figure}[t!]
     \centering
     \begin{subfigure}[t]{\columnwidth}
          \centering
             \begin{tikzpicture}
  \begin{axis}[
   width=12cm,
    height=4cm,
    ymin=0,
   xticklabels={},
   yticklabel={\pgfmathparse{\tick*100}\pgfmathprintnumber{\pgfmathresult}\%},
    ylabel={CV of $\ell$},
      axis x line  = bottom,
    axis y line  = left  ,
    legend cell align={left},
    legend columns=3,
    legend style={font=\footnotesize,draw=none,fill=none,at={(0.7,1.2)}},             
  ]

\addplot [mark=o,blue!70] plot coordinates {
(0, 0.18060750261208017)
(1, 0.18729768067485847)
(2, 0.2625158784963267)
(3, 0.25474533739007826)
(4, 0.1460048780565191)
(5, 0.37231217419714446)
(6, 0.29962460536081964)
(7, 0.3029131476123129)
(8, 0.43502908238472165)
(9, 0.31407377348583976)
(10, 0.300699599575852)
(11, 0.39107542733457085)
(12, 0.36909045425995063)
(13, 0.3174496317585824)
(14, 0.2775972728831085)
(15, 0.5125528425658595)
(16, 0.37120563640298326)
(17, 0.2590466451121906)
(18, 0.45927827449164493)
(19, 0.308405540476585)
(20, 0.3508425232356721)
(21, 0.2904621997209649)
(22, 0.38566709998218834)
(23, 0.29995853664432764)
(24, 0.3306926574916014)
(25, 0.30110336212994776)
(26, 0.3479099103200286)
(27, 0.27609264035655073)
(28, 0.25448480816650554)
(29, 0.2249598926758904)
(30, 0.30990130736188787)
(31, 0.2738344325319285)
(32, 0.28204373247612785)
(33, 0.251171258281552)
(34, 0.30920704744144994)
(35, 0.3529107622725212)
(36, 0.24658477743769142)
(37, 0.21720153485329588)
(38, 0.3487525992523332) 
(39, 0.27168611839474804)
(40, 0.37310407726890865) 
(41, 0.2208098915192508)
(42, 0.25507139703137754)
(43, 0.28695546642314246) 
(44, 0.36861947361265357)
(45, 0.20769982269139986)
(46, 0.5142966362021985)
(47, 0.43273809726243495)
(48, 0.24897572903132348)
(49, 0.37753309235557136)
(50, 0.22397953099921317)
(51, 0.36929973575536096)
(52, 0.25779044009030105)
(53, 0.396817269846218)
(54, 0.26438671355142596)
(55, 0.38239438103251594)
(56, 0.22262773541417552)
(57, 0.1508872630839325)
(58, 0.40904127876931096)
(59, 0.3711678953176816)
};

\addplot [mark=+,red!70] plot coordinates {
(0, 0.3352465704829911)
(1, 0.29882814518602685)
(2, 0.2790215386747226)
(3, 0.36816180017759614)
(4, 0.31619560634602367)
(5, 0.3627647203068042) 
(6, 0.2898448264102677)
(7, 0.3089234755433479)
(8, 0.27608387324664946)
(9, 0.23945611749133314)
(10, 0.2794364845340401) 
(11, 0.22875142278596058) 
(12, 0.3144052681995326)
(13, 0.13337080350561414)
(14, 0.2521577588992209)
(15, 0.3158780975252126)
(16, 0.2578257001853256)
(17, 0.3041207723396091)
(18, 0.27740109204201124)
(19, 0.2488091729894105)
(20, 0.23674596967859282)
(21, 0.3410002997413745)
(22, 0.28612468111628403)
(23, 0.29573947280233925)
(24, 0.258315649215395)
(25, 0.2939676541668615)
(26, 0.3861337201492284)
(27, 0.43139590451541965)
(28, 0.28270516396987155)
(29, 0.2492776711950993) 
(30, 0.20343466595201998)
(31, 0.3895912507533517)
(32, 0.26759341355185484)
(33, 0.1823627739030176)
(34, 0.36252437192122544)
(35, 0.224468371387495)
(36, 0.3332324494556924)
(37, 0.4438166017733331)
(38, 0.22669049891106985)
(39, 0.20631961167686289)
(40, 0.41131868731412713)
(41, 0.2530777921223463)
(42, 0.1663011966261366) 
(43, 0.4291224685988321)
(44, 0.1925522980645313)
(45, 0.2874451900777021)
(46, 0.26077511301577083)
(47, 0.2766493495634544)
(48, 0.3096332470615113)
(49, 0.3827831532223824)
(50, 0.326907656741242)
(51, 0.25884678214202117)
(52, 0.3182283020365298)
(53, 0.2789963149782796)
(54, 0.3988013414669605)
(55, 0.23124247527161818)
(56, 0.19601953283156018)
(57, 0.22164516421363684)
(58, 0.544366708115364)
(59, 0.35852967916123346)
};

\addplot [mark=x,teal!70] plot coordinates {
(0, 0.2122702770346339)
(1, 0.35886393908417025) 
(2, 0.295658257044015)
(3, 0.36680209124996316)
(4, 0.2228000614205177)
(5, 0.3762606912907723)
(6, 0.21633620934658052)
(7, 0.3978902399525025)
(8, 0.2655153869113546)
(9, 0.3392010165841507)
(10, 0.24097688697305328)
(11, 0.3452143536736174)
(12, 0.16488097526246556)
(13, 0.30059166479954946)
(14, 0.3965226668314706)
(15, 0.31978780424255676)
(16, 0.3406522846462826)
(17, 0.34139624238278776)
(18, 0.3380120945594194)
(19, 0.27655706046381556)
(20, 0.28081898556709534)
(21, 0.27504280052789193)
(22, 0.2650208602579497)
(23, 0.2656470395731377)
(24, 0.396533977774384)
(25, 0.14958812394287369)
(26, 0.2148659464004434)
(27, 0.36192422950460573)
(28, 0.2960496173167294)
(29, 0.2865717796747426)
(30, 0.36599236693380743)
(31, 0.2786694667169542)
(32, 0.3444800180426513)
(33, 0.27520184186398167)
(34, 0.5114872845792083)
(35, 0.29377170445572115)
(36, 0.3017783060195155)
(37, 0.2636084136618488)
(38, 0.2790908514277102)
(39, 0.26545327515728473)
(40, 0.3089200529570269)
(41, 0.37816969854788374) 
(42, 0.23945316492254598)
(43, 0.27428510732974165)
(44, 0.2960977399188241) 
(45, 0.4176928816179101)
(46, 0.39237420810455176)
(47, 0.3259762346042035) 
(48, 0.26925456666320735)
(49, 0.36231543523536724)
(50, 0.16039857469843583)
(51, 0.21025941013407676)
(52, 0.3517989203775896)
(53, 0.2461415582827543)
(54, 0.3741546314447714)
(55, 0.3003500802352634)
(56, 0.34301509948454934)
(57, 0.4022895460691054)
(58, 0.3377173658789333)
(59, 0.24114754755628595)
};

  \addplot [dashed,black] plot coordinates {
(0,0.05)
(59,0.05)
  };

  \legend{
HP,
SB,
LB
};


  \end{axis}

\end{tikzpicture}
     \end{subfigure}
     \vspace{-0.2cm}
     \begin{subfigure}[t]{\columnwidth}
          \centering
             \begin{tikzpicture}
  \begin{axis}[
   width=12cm,
    height=4cm,
    ymin=0,
   xticklabels={},
   yticklabel={\pgfmathparse{\tick*100}\pgfmathprintnumber{\pgfmathresult}\%},
    ylabel={CV of $\varrho$},
    xlabel={Combination of metric and dataset.},
      axis x line  = bottom,
    axis y line  = left  ,
    legend cell align={left},
    legend columns=3,
    legend style={font=\footnotesize,draw=none,fill=none,at={(0.6,1.2)}},             
  ]

\addplot [mark=o,blue!70] plot coordinates {
(0, 0.5123575335720734)
(1, 0.7273952002335032)
(2, 0.3862555392789808)
(3, 0.34685541595812186)
(4, 0.3120774219226788)
(5, 0.31419849616696344)
(6, 0.5864132996592972)
(7, 0.5570465719533788)
(8, 0.3285505131056037)
(9, 0.5174550260304741)
(10, 0.41554157316149665)
(11, 0.3156210552784179)
(12, 0.4581774415669264)
(13, 0.31499660478651714)
(14, 0.2834051838440178)
(15, 0.4666317292135073)
(16, 0.27149033573502046)
(17, 0.29081353018885353)
(18, 0.3533062189942781)
(19, 0.23791142907920154)
(20, 0.23421835802587282)
(21, 0.28996642930979993)
(22, 0.40640466798523434)
(23, 0.09454848525246641)
(24, 0.35667907937265186)
(25, 0.8214539190958265)
(26, 0.6652763946823944)
(27, 0.4748319191305026)
(28, 0.43219455198454926)
(29, 0.47197377299301496)
(30, 0.38181505878467115)
(31, 0.2795560617762159)
(32, 0.5152606957584149)
(33, 0.381418181607623)
(34, 0.5481765643274498)
(35, 0.4402909158250129)
(36, 0.40914531707926677)
(37, 0.4321981810458263)
(38, 0.284198840642121)
(39, 0.49220447784981264)
(40, 0.4294277069777396)
(41, 0.44263485529899566)
(42, 0.47577736614471167)
(43, 0.9187465968963102)
(44, 0.23148801085411225)
(45, 0.3492534102765623)
(46, 0.31475083257313197)
(47, 0.2301717434425675)
(48, 0.46974000899020574)
(49, 0.46383549039274946)
(50, 0.7682464360473)
(51, 0.40004137575510473)
(52, 0.38509019568307956)
(53, 0.3623504389883517)
(54, 0.28557327226909746)
(55, 0.5745107162520782)
(56, 0.7236157996112921)
(57, 0.2127063025438728)
(58, 0.45921885540047364)
(59, 0.5550542714989993)
};

\addplot [mark=+,red!70] plot coordinates {
(0, 0.6174327078319484)
(1, 0.4163848334319148)
(2, 0.7581559534908078)
(3, 0.3893274131034796)
(4, 0.37589626505132057)
(5, 0.5268045813606672)
(6, 0.1897507746997584)
(7, 0.4620966347993007)
(8, 0.5079731080417941)
(9, 0.6312277988375772)
(10, 0.5823219028270812)
(11, 0.5760490539066221)
(12, 0.5107122890615972)
(13, 0.5430883622772114)
(14, 0.33933991087267024)
(15, 0.3457061024403412)
(16, 0.2570837530081654)
(17, 0.6373855076570862)
(18, 0.4743744295345764)
(19, 0.8040664739070913)
(20, 0.3342009988358853)
(21, 0.3375489826840049)
(22, 0.6767827367341425)
(23, 0.3809091246944537)
(24, 0.4315268118309244)
(25, 0.4010333784343841)
(26, 0.4183893448839545)
(27, 0.547301176275814)
(28, 0.6184967851862024)
(29, 0.48596124487612585)
(30, 0.34695842130973004)
(31, 0.3606224632023702)
(32, 0.3856786271639843)
(33, 0.4287264784534992)
(34, 0.5206819348399006)
(35, 0.5273356810171026)
(36, 0.6902588373166476)
(37, 0.4318548543496853)
(38, 0.5030254481571427)
(39, 0.8175562502011693)
(40, 0.4879325397510705)
(41, 0.3331320632008804)
(42, 0.37792641375455255)
(43, 0.4480981728813022)
(44, 0.4082924113442226)
(45, 0.34562136826873086)
(46, 0.3157812477429478)
(47, 0.6290755865891917)
(48, 0.844943479833499)
(49, 0.2562973202006251)
(50, 0.4017433515880175)
(51, 0.7557025154526333)
(52, 0.623631534416487)
(53, 0.4814260990489072)
(54, 0.24657233401922485)
(55, 0.5205794226023036)
(56, 0.5879135749903953)
(57, 0.5370249064656789)
(58, 0.3503002545237464)
(59, 0.3068101855783595)
};

\addplot [mark=x,teal!70] plot coordinates {
(0, 0.7078588988131365)
(1, 0.4096424372393714) 
(2, 0.1399485530016954)
(3, 0.42067342395020213)
(4, 0.7600921379875301)
(5, 0.40942282813374536)
(6, 0.646991064716807)
(7, 0.4389843067109109)
(8, 0.3368891877801953)
(9, 0.6190838624463304)
(10, 0.26859281830918)
(11, 0.4323873602166993)
(12, 0.46547883723880434)
(13, 0.6985644865478303)
(14, 0.3292874860670651)
(15, 0.6662475884341297)
(16, 0.6887834033472859)
(17, 0.3444415196065909)
(18, 0.3094802849771597)
(19, 0.6016913048828443)
(20, 0.30177497817087423)
(21, 0.4779501300584445)
(22, 0.5488359201354098)
(23, 0.37694847794971864)
(24, 0.49415404780751493)
(25, 0.5161548367808794)
(26, 0.284624076901746)
(27, 0.5585131004969672)
(28, 0.28844173642645116)
(29, 0.40956167513214475)
(30, 0.35325115898555765)
(31, 0.577860550159416)
(32, 0.4829280734322184)
(33, 0.3818764527184661)
(34, 0.5621403983342562)
(35, 0.4510538548159614)
(36, 0.3966086959724363)
(37, 0.19099048909302924)
(38, 0.6921698693320006)
(39, 0.4351082260836895)
(40, 0.722008017960146)
(41, 0.34631415254573833)
(42, 0.5518347584831758)
(43, 0.26196832526077035)
(44, 0.5718145356236176)
(45, 0.7784164100836723)
(46, 0.9892536392713984)
(47, 0.5724293833911599)
(48, 0.49348877876717473)
(49, 0.7731038671721637)
(50, 0.4173668487405264)
(51, 0.35280232763989017)
(52, 0.2483558961532205)
(53, 0.7936566447885994)
(54, 0.5350540669524265)
(55, 0.35225210574137444)
(56, 0.5257673867606381)
(57, 0.6250603569126949)
(58, 0.21461770653434217)
(59, 0.27219423656355013)
};

  \addplot [dashed,black] plot coordinates {
(0,0.05)
(59,0.05)
  };



  \end{axis}

\end{tikzpicture}
     \end{subfigure}
      \vspace{-0.2cm}
      \caption{CV of FDC and correlation length on metric-dataset pairs. The dashed line indicates the threshold of $5\%$ CV.}
    \label{fig:rq4-2}
    \vspace{-0.5cm}
\end{figure}
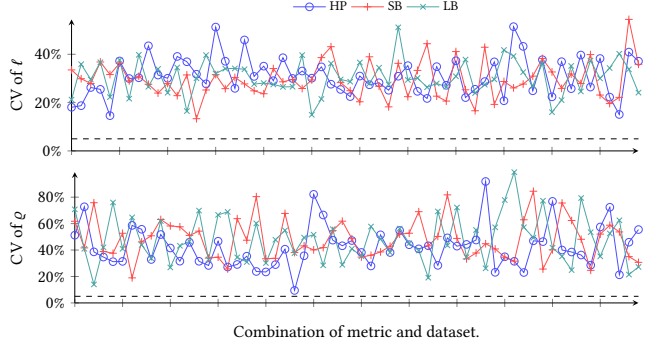

Likewise, Figure~\ref{fig:rq4-2} illustrates the CV for each of the 60 cases of changing context settings. Software engineering researchers often use $5\%$ as a threshold of whether the variability is significant~\cite{DBLP:conf/kbse/MalavoltaGLVTZP20,DBLP:conf/icse/WeberKSAS23,DBLP:journals/toit/LeitnerC16}. That is, $CV>5\%$ implies that there is likely to be a substantial fluctuation. With CV, we aim to examine whether the changing context settings would considerably affect the difficulty of the testing landscape. As can be seen, all the cases show $CV>5\%$ across the three context types; in most of the cases, it is greater than $50\%$, suggesting a drastic shift in the testing landscape. This explains why in RQ$_2$ that varying context settings would generally lead to significant changes in the test adequacy results.

\begin{quotebox}
   \noindent
   \textit{\textbf{Response to RQ$_3$:}  On test adequacy, compared with the baseline, non-optimized hyperparameters reduce the guidance, making the testing landscape harder to cover. Selection and label bias provide stronger guidance while rendering the landscape smoother, relaxing the difficulty of testing. The changing context settings lead to considerable variation in the structure of the testing landscape.}
\end{quotebox}

\begin{figure}[t!]
\centering
  \begin{subfigure}[t]{\columnwidth}
          \centering
          \includegraphics[width=\columnwidth]{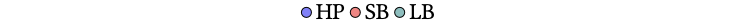}
     \end{subfigure}
       \vspace{-0.3cm}
{
    
     \begin{subfigure}[t]{0.3\columnwidth}
          \centering
          \includegraphics[width=\columnwidth]{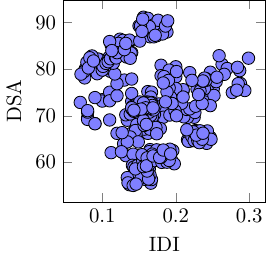}
          \vspace{-0.6cm}
          \subcaption*{($r=-0.14, p<0.01$)}
     \end{subfigure}
     ~
        \begin{subfigure}[t]{0.3\columnwidth}
          \centering
          \includegraphics[width=\columnwidth]{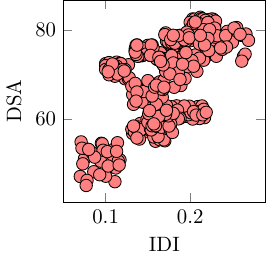}
         \vspace{-0.6cm}
          \subcaption*{($r=0.58, p<0.01$)}
     \end{subfigure}
     ~
         \begin{subfigure}[t]{0.3\columnwidth}
          \centering
          \includegraphics[width=\columnwidth]{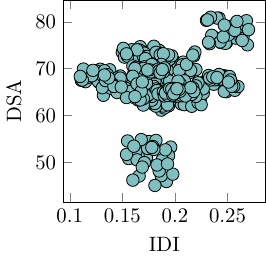}
          \vspace{-0.6cm}
          \subcaption*{($r=0.10, p=0.054$)}
     \end{subfigure}
      \vspace{-0.2cm}
\subcaption{DSA-IDI}

    \begin{subfigure}[t]{0.29\columnwidth}
          \centering
          \includegraphics[width=\columnwidth]{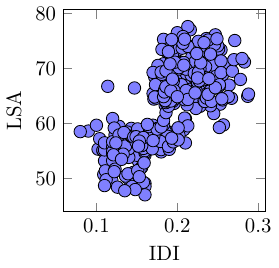}
          \vspace{-0.6cm}
          \subcaption*{($r=0.72, p<0.01$)}
     \end{subfigure}
     ~
        \begin{subfigure}[t]{0.3\columnwidth}
          \centering
          \includegraphics[width=\columnwidth]{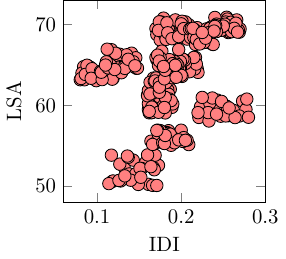}
          \vspace{-0.6cm}
          \subcaption*{($r=0.35, p<0.01$)}
     \end{subfigure}
     ~
         \begin{subfigure}[t]{0.28\columnwidth}
          \centering
          \includegraphics[width=\columnwidth]{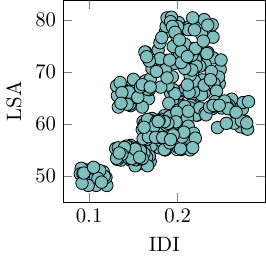}
          \vspace{-0.6cm}
          \subcaption*{($r=0.59, p<0.01$)}
     \end{subfigure}
       \vspace{-0.2cm}
\subcaption{LSA-IDI}

    \begin{subfigure}[t]{0.29\columnwidth}
          \centering
          \includegraphics[width=\columnwidth]{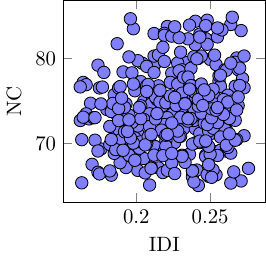}
          \vspace{-0.6cm}
          \subcaption*{($r=0.11, p=0.031$)}
     \end{subfigure}
     ~
        \begin{subfigure}[t]{0.3\columnwidth}
          \centering
          \includegraphics[width=\columnwidth]{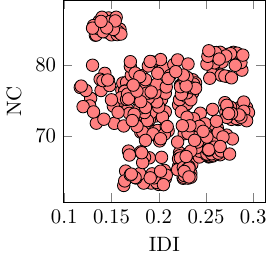}
          \vspace{-0.6cm}
          \subcaption*{($r=-0.22, p<0.01$)}
     \end{subfigure}
     ~
         \begin{subfigure}[t]{0.29\columnwidth}
          \centering
          \includegraphics[width=\columnwidth]{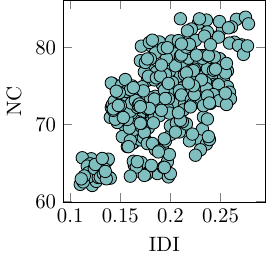}
          \vspace{-0.6cm}
          \subcaption*{($r=0.59, p<0.01$)}
     \end{subfigure}
       \vspace{-0.2cm}
\subcaption{NC-IDI}

    \begin{subfigure}[t]{0.29\columnwidth}
          \centering
          \includegraphics[width=\columnwidth]{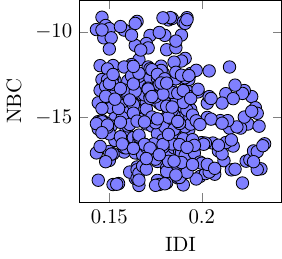}
          \vspace{-0.6cm}
          \subcaption*{($r=-0.20, p<0.01$)}
     \end{subfigure}
     ~
        \begin{subfigure}[t]{0.29\columnwidth}
          \centering
          \includegraphics[width=\columnwidth]{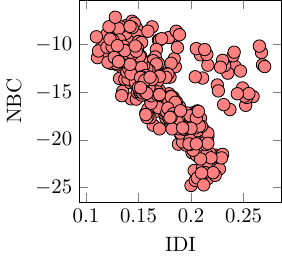}
          \vspace{-0.6cm}
          \subcaption*{($r=-0.66, p<0.01$)}
     \end{subfigure}
     ~
         \begin{subfigure}[t]{0.31\columnwidth}
          \centering
          \includegraphics[width=\columnwidth]{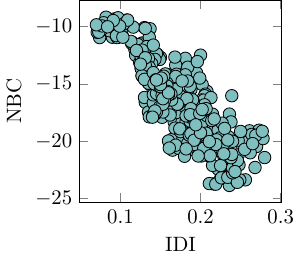}
          \vspace{-0.6cm}
          \subcaption*{($r=-0.85, p<0.01$)}
     \end{subfigure}
       \vspace{-0.2cm}
\subcaption{NBC-IDI}

    \begin{subfigure}[t]{0.3\columnwidth}
          \centering
          \includegraphics[width=\columnwidth]{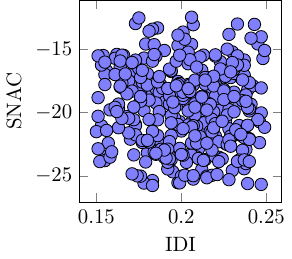}
          \vspace{-0.6cm}
          \subcaption*{($r=-0.05, p=0.39$)}
     \end{subfigure}
     ~
        \begin{subfigure}[t]{0.29\columnwidth}
          \centering
          \includegraphics[width=\columnwidth]{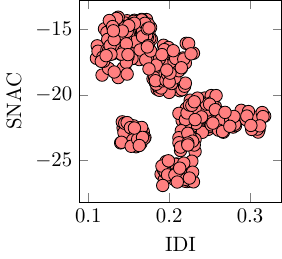}
          \vspace{-0.6cm}
          \subcaption*{($r=-0.57, p<0.01$)}
     \end{subfigure}
     ~
         \begin{subfigure}[t]{0.29\columnwidth}
          \centering
          \includegraphics[width=\columnwidth]{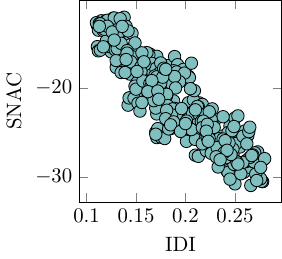}
          \vspace{-0.6cm}
          \subcaption*{($r=-0.89, p<0.01$)}
     \end{subfigure}
       \vspace{-0.2cm}
\subcaption{SNAC-IDI}
     }
         \vspace{-0.3cm}
    \caption{Spearman correlations (and their $p$ values) between adequacy and fairness by contexts. (NBC and SNAC are converted as maximizing for the convenience of interpretation).}
    \label{fig:sp}
\end{figure}

\subsection{Correlation of Adequacy and Fairness (RQ$_4$)}

\subsubsection{Method}

We leverage Spearman correlation ($r$)~\cite{myers2004spearman}, which is a widely used metric in software engineering~\cite{DBLP:conf/icse/Chen19b,DBLP:journals/tse/WattanakriengkraiWKTTIM23}, to quantify the relationship between a given adequacy metric and the fairness metric. Specifically, Spearman correlation measures the nonlinear monotonic relation between two random variables and we have $-1\leq r \leq1$. $r$ represents the strength of monotonic correlation and $r=0$ means that the two variables do not correlate with each other in any way; $-1\leq r<0$ and $0< r \leq 1$ denote that the correlation is negative and positive, respectively. For each pair of adequacy and fairness metrics, we calculate the data points for all datasets, context types/settings, generators, and runs.

To interpret the strength of Spearman correlation, we follow the common pattern below, which has also been widely used in software engineering~\cite{DBLP:journals/infsof/SamoladasAS10,DBLP:journals/tse/WattanakriengkraiWKTTIM23}: $0 \leq |r| \leq 0.09$ is negligible; $0.09 < |r| \leq 0.39$ implies weak; $0.39 < |r| \leq 0.69$ is considered as moderate; $0.69 < |r| \leq 0.89$ is strong, and $0.89 < |r| \leq 1$ means very strong.


\subsubsection{Result}

In Figure~\ref{fig:sp}, we plot the distribution and Spearman correlation on all data points for each pair of adequacy and fairness metric. All the adequacy metrics (including the converted NBC and SNAC) are to be maximized, therefore $r>0$ means that the adequacy is indeed helpful for improving the fairness metric; $r<0$ implies that the adequacy metrics could turn out to be misleading. We also report the $p$-values to verify if $r$ statistically differs from $0$.

We note that DSA, LSA, and NC are generally helpful while NBC and SNAC tend to be misleading. In terms of the context type, we see that each of them exhibits different correlations depending on the adequacy metrics. Surprisingly, different hyperparameters often lead to weak monotonic correlations (except for LSA) while diverse selection and label bias most commonly have moderate to strong monotonic correlations for both the positive and negative $r$. 

The above explains the reason behind the discrepancy between the results of adequacy and fairness metrics we found for RQ$_1$ and RQ$_2$: for changing hyperparameters, the weak correlations mean that test adequacy is often neither useful nor harmful, or at least only marginally influential, for improving the fairness metric monotonically. For varying selection and label bias, in contrast, the effectiveness of those adequacy metrics that are highly useful for fairness (e.g., DSA and LSA) has been obscured by the results of those that can mislead the fairness results, e.g., NBC and SNAC. 


\begin{quotebox}
   \noindent
   \textit{\textbf{Response to RQ$_4$:} For changing hyperparameters, the test adequacy often has a weak monotonic correlation to the fairness metrics. However, when varying selection and label bias, adequacy can strongly help or mislead the fairness metric, depending on the test adequacy metric used.}
\end{quotebox}
\section{Discussion on Insights}
\label{sec:discussions}



The findings above shed light on the strategies to design and evaluate a test generator for fairness testing, which we discuss here.



Contrary to existing work that often only assumes an ideal and fixed context setting in the evaluation~\cite{DBLP:journals/corr/abs-1905-05786,chakraborty2020making,gohar2023towards,tizpaz2022fairness,chen2022maat,chakraborty2020fairway,chakraborty2021bias,DBLP:conf/wacv/KarkkainenJ21,mambreyan2022dataset,zhang2021ignorance,DBLP:conf/fairware-ws/ChakrabortyMT22,DBLP:journals/corr/abs-2108-08504}, from RQ$_1$ and RQ$_2$, a key insight we obtained is that the context types often have considerable implications for the test results. In particular, different types of context can influence the testing differently, i.e., compared with the perfect baseline, the hyperparameter influences the results in a completely different manner against that for selection and label bias. Varying the concrete setting can also lead to significantly different results and hence pose an additional threat to external validity. We would like to stress that, we do not suggest abandoning any methods used for setting the context, e.g., hyperparameter tuning. Our results simply challenge the current practice that fairness testing is evaluated based on a fixed setting, e.g., well-tuned hyperparameters only, which might not always be possible given the expensive tuning process, e.g., the hyperparameters might be sub-optimal in practice. All those observations are crucial for properly evaluating the test generators in future work.




\begin{quotebox}
   \noindent
       \textit{\textbf{Insight 1:} Testing the fairness of deep learning systems at the model level must evaluate diverse settings of the contexts. Notably, we should test the fairness of deep learning systems beyond a ``perfect'' condition, as sub-optimal hyperparameters (which is not uncommon in practice) can make the generators more struggle while the presence of data bias often boosts the performance of generators.}
\end{quotebox}



Existing test generators for fairness testing are mostly designed based on the understanding of the DNN architecture~\cite{udeshi2018automated,zhang2020white,aggarwal2019black,galhotra2017fairness,aggarwal2019black,DBLP:conf/sigsoft/TaoSHF022,DBLP:conf/issta/ZhangZZ21,DBLP:conf/icse/ZhengCD0CJW0C22}, while our results from RQ$_3$ provide new insights for designing them from the perspective of the testing landscape. For example, a higher FDC on test adequacy means that it is less reliable to use the distance between test cases as an indication of their tendency of fitness change during mutation operation.




\begin{quotebox}
   \noindent
   \textit{\textbf{Insight 2:} For better test adequacy with changing hyperparameter settings, current generators should be more capable of dealing with ineffective guidance in the testing landscape. Using existing generators to test under diverse conditions of selection/label bias can focus more on the speed of improving adequacy, thanks to the easier landscape structure.}
\end{quotebox}



While researchers believe that better test adequacy would most likely help find fairness bugs~\cite{DBLP:journals/tosem/ChenWMSSZC23,DBLP:journals/corr/abs-2207-10223}, our results from $RQ_4$ show that, depending on the context type, this is not always true. Specifically, we demonstrate the in-depth correlation between test adequacy and fairness metrics with respect to the context, revealing that improving test adequacy does not necessarily lead to better fairness bug discovery; rather, it can be misleading. This does not only provide insights for future research, e.g., to conduct casual analysis behind the correlations but also directs the test generator design since following test adequacy may not always be reliable.


\begin{quotebox}
   \noindent
   \textit{\textbf{Insight 3:} Changing hyperparameter settings often leads to a weak correlation between the test adequacy and fairness metrics, and therefore generators do not have to strictly follow the adequacy, i.e., lower adequacy can be accepted without harming fairness bug discovery. For varying selection and label bias, metrics that focus on the boundary of the neurons like NBC and SNAC should be avoided, as they are less likely to be beneficial for finding fairness bugs. }
\end{quotebox}
\section{Threats to validity}
\label{sec:tov}


\textbf{Internal threats} may be related to the selected contexts, their settings, and test generators. To mitigate this, we conducted a systematic review to identify the most common options and follow the settings from existing work. To avoid bias in the sampling of the context settings, we adopt Latin Hypercube sampling~\cite{stein1987large} while keeping the overhead acceptable. Indeed, all three generators studied are white-box that penetrate into the internal aspects of DNN. This, in contrast to their black-box counterpart, is more suitable for our case since we know what type of DNN to be tested as confirmed by existing work~\cite{DBLP:journals/corr/abs-2207-10223,DBLP:journals/corr/abs-2111-08856}. We anticipate that our conclusions can also be well reflected in the black box method since they can be more sensitive to the context change (because they exploit even less information about the model to be tested). However, unintended omission of information or options is always possible.


\textbf{Construct threats} on validity may be related to the metrics used. We select five test adequacy metrics and one fairness metric, due to their popularity, category, and the requirements of our study setup, e.g., targeting individual fairness. To understand the causes of the observations on testing results, we leverage common metrics from fitness landscape analysis at both the local and global levels, revealing information from different aspects. To ensure statistical significance, the Scott-Knott test~\cite{DBLP:journals/tse/MittasA13}, Wilcoxon sum-rank test~\cite{mann1947test}, and Harmonic mean $p$-values~\cite{wilson2019harmonic} are used where required. Indeed, a more exhaustive study of metrics can be part of our future work.


Finally, \textbf{external threats} to validity can come from the generalizability concern. To mitigate this, we consider 12 datasets for deep learning-based systems, three context types with 10 settings each, and three test generators, leading to $10,800$ cases. Such a setup, while not exhaustive, serves as a solid basis for generalizing our findings considering the limited resources. Yet, we agree that examining more diverse subjects may prove fruitful.


\section{Related Work}
\label{sec:related}




\textbf{Test Cases Generation for Fairness:} Many test generators have been proposed for fairness testing at the model level~\cite{udeshi2018automated,zhang2020white,aggarwal2019black,galhotra2017fairness,aggarwal2019black,DBLP:conf/sigsoft/TaoSHF022,DBLP:conf/issta/ZhangZZ21,DBLP:conf/icse/ZhengCD0CJW0C22}. Among others, Aggarwal \textit{et al.}~\cite{aggarwal2019black} propose a black-box test generator that uses symbolic generation for individual fairness of classification-based models in learning systems. Zhang \textit{et al.}~\cite{zhang2020white} present a white-box generator to detect individual discriminatory instances. Through gradient computation and clustering, the generator performs significantly more scalable than existing methods. More recently, Zheng \textit{et al.}~\cite{DBLP:conf/icse/ZhengCD0CJW0C22} present a generator for fairness testing by leveraging identified biased neurons. It detects the biased neurons responsible for causing discrimination via neuron analysis and generates discriminatory instances, which is achieved by optimizing the objective of amplifying the activation difference values of these biased neurons. Similarly, Tao \textit{et al.}~\cite{DBLP:conf/sigsoft/TaoSHF022} propose a generator that uses perturbation restrictions on sensitive/non-sensitive attributes for generating test cases at the model level. Yet, those generators commonly assume fixed conditions from the other components in deep learning systems, leaving the validity of the conclusion in a different context questionable.




\textbf{Fairness Bugs Mitigation:} Much work exists that uses the test cases to improve fairness~\cite{zhang2021ignorance,chakraborty2020fairway,DBLP:conf/fairware-ws/ChakrabortyMT22}. For example, Zhang and Harman~\cite{zhang2021ignorance} explore the factors that affect model fairness and propose a tool that leverages enlarging the feature set as a possible way to improve fairness. Chakraborty \textit{et al.}.~\cite{chakraborty2020fairway} remove ambiguous data points in training data and then apply multi-objective optimization to train fair models. A follow-up work of Chakraborty \textit{et al.}~\cite{DBLP:conf/fairware-ws/ChakrabortyMT22} not only removes ambiguous data points but also balances the internal distribution of training data. Nevertheless, fairness issue mitigation is often isolated from fairness testing.

\textbf{Relevant Empirical Studies:} Empirical studies have also been conducted on testing deep learning systems or on the fairness-related aspects. Biswas and Rajan~\cite{biswas2020machine} perform an empirical study examining a variety of fairness improvement techniques. Hort \textit{et al.}~\cite{hort2021fairea} conduct a similar study but with a larger scale. Recently, Chen \textit{et al.}~\cite{chen2023comprehensive} provide a study of 17 representative fairness repair methods to explore their influences on performance and fairness. The above are relevant to fixing fairness bugs rather than testing them. Prior study has also shown that NC might not be a useful test adequacy metric when guiding the robustness testing of model accuracy~\cite{DBLP:conf/sigsoft/Harel-CanadaWGG20}. Yet, the results cannot be generalized to fairness metrics and are irrelevant to the context setting of fairness testing. Zhang and Harman~\cite{DBLP:conf/icse/ZhangH21} present an empirical study on how the feature dimension and size of training data affect fairness. However, they have not covered the ability of test generators to find fairness bugs therein. Recently, there has been a study that explores the correlation between test adequacy and fairness metrics~\cite{DBLP:journals/tse/ZhengLWC24}, from which the authors partially confirmed our results from \textbf{RQ$_3$}. However, that study does not consider context types and includes only a much smaller set of adequacy metrics than our work, but it serves as evidence of the strong interest in the community in using test adequacy metrics to guide fairness testing.

Unlike the above, our work explicitly studies the implication of contexts on fairness testing at the model level, from which we reveal important findings and insights that were previously unexplored.
\section{Conclusion}
\label{sec:conclusion}

This paper fills the gap in understanding \textit{how} and \textit{why} the context from the other parts of a deep learning system can influence the fairness testing at the model level. We do so with 12 datasets for deep learning systems, three context types with 10 settings each, three test generators, and five test adequacy metrics, leading to $10,800$ cases of investigation---the largest scale study on this matter to the best of our knowledge. We reveal that:

\begin{itemize}
  \item Distinct context types can influence the existing test generators for fitness testing at the model level in opposed ways while the settings also create significant implications.
  \item The observations are due to the change in landscape structure, particularly on the search guidance and ruggedness.
  \item The correlation between test adequacy metrics is not always reliable in improving the discovery of fairness bugs. 
\end{itemize}

We articulate a few actionable insights for future research on fairness testing at the model level:
\begin{itemize}
    \item It is important to consider context types and their settings when evaluating current test generators.
    \item To improve existing generators on non-optimized hyperparameters for test adequacy, one would need to focus on mitigating ineffective fitness guidance. For current generators with different levels of data bias, the focus can be on improving the efficiency with the test adequacy metric.
    \item While test adequacy metrics can be helpful in discovering fairness bugs, they might not always be useful. Sometimes, it can be beneficial to not leverage them at all.
\end{itemize}

We hope that our findings bring attention to the importance of contexts for fairness testing on deep learning systems at the model level, sparking the dialogue on a range of related future research directions in this area.




\begin{acks}
This work was supported by a UKRI Grant (10054084) and a NSFC Grant (62372084).
\end{acks}

\balance
\bibliographystyle{ACM-Reference-Format}
\bibliography{references}

\end{document}